\documentclass[10pt]{article}

\usepackage{ctable}
\usepackage{tabularx}
\usepackage{graphicx}
\usepackage{adjustbox}
\usepackage[percent]{overpic}
\usepackage[font={small}]{caption}

\begin{document}

\title{{\it VoroTop}: Voronoi Cell Topology \\Visualization and Analysis Toolkit}
\author{Emanuel A. Lazar\thanks{University of Pennsylvania, Philadelphia, PA 19104, mLazar$@$seas.upenn.edu}}
\date{\today}
\maketitle

\begin{abstract}
This paper introduces a new open-source software program called {\it VoroTop}, which uses Voronoi topology to analyze local structure in atomic systems.  Strengths of this approach include its abilities to analyze high-temperature systems and to characterize complex structure such as grain boundaries.  This approach enables the automated analysis of systems and mechanisms previously not possible.
\end{abstract}

\section{Introduction}
A central challenge of studying condensed matter systems on the atomic scale is the development of a meaningful way to analyze structure, so that identification of defects and characterization of complex structure can be performed in an automated manner.  The analysis of perfect crystals, both in two and three dimensions, is greatly aided by mathematical tools such as group theory, which help translate fuzzy ideas about order, structure, and symmetry into mathematically rigorous language.  In turn, mathematical analysis provides physical understanding of many physical systems.  Loosely speaking, the structure of a crystal can be completely specified by a set of linearly independent vectors whose integral combinations specify coordinates of all atoms (i.e., a basis).  Defects in perfect crystals can be readily identified as deviations from this ideal structure.  For example, a vacancy can be understood as a lattice site missing an atom, and an interstitial can be understood as an atom located at a non-lattice site.  More complex defect structures can be viewed along similar lines.

This approach towards structure analysis is insufficient for understanding real systems.  Consider, for example, the simple case of a finite-temperature crystal, in which all atoms are slightly perturbed from their lattice sites.  The conventional approach might lead us to describe such a system as being crystalline but populated by numerous vacancies and interstitials.  This absurdity is inconsistent with our intuition, and motivates the development of a more robust approach towards characterizing structure in atomic systems.  

Many approaches to address this problem have been developed in recent decades \cite{stukowski2012structure}.  Most of these approaches describe each arrangement of neighboring atoms by quantifying how similar or different it is from an ideal reference configuration.  Although such methods are well-suited for studying low-temperature systems, they perform poorly when applied to systems at temperatures above half of their bulk melting temperatures, or systems otherwise strongly perturbed from their ground state \cite{stukowski2012structure, lazar2015topological}.  In recent work we have explained some fundamental limitations of many conventional approaches \cite{landweber2016fiber} and described a more robust one based on Voronoi topology \cite{lazar2015topological}.  

This paper introduces an open-source software program called {\it VoroTop} that automates this analysis.  The name of the software derives from the words `Voronoi topology', the theory on which it is based.  {\it VoroTop} makes this analysis accessible to a broad audience through a command-line program and library.  This paper is organized as follows.  Section \ref{basics} describes the basics of Voronoi cells, and explains how their topology can be used to analyze structure in atomic systems.  Section \ref{extensions} describes several extensions of the basic theory.  Section \ref{compare} compares this approach with several conventional ones, and Section \ref{case} illustrates the utility of {\it VoroTop} in visualizing and analyzing the behavior of two screw dislocations moving inside a high-temperature crystal.  Section \ref{software} describes the {\it VoroTop} software itself and its core functions and features.

\section{Voronoi topology basics}
\label{basics}

\subsection{Voronoi cells and local structure}
\label{vctls}

\begin{figure}
\setlength{\tabcolsep}{2.2pt}
\begin{center}
\begin{tabular}{ccc}
\frame{\includegraphics[width=0.32\columnwidth,trim={0.cm 3.55cm 0.cm 3.55cm},clip]{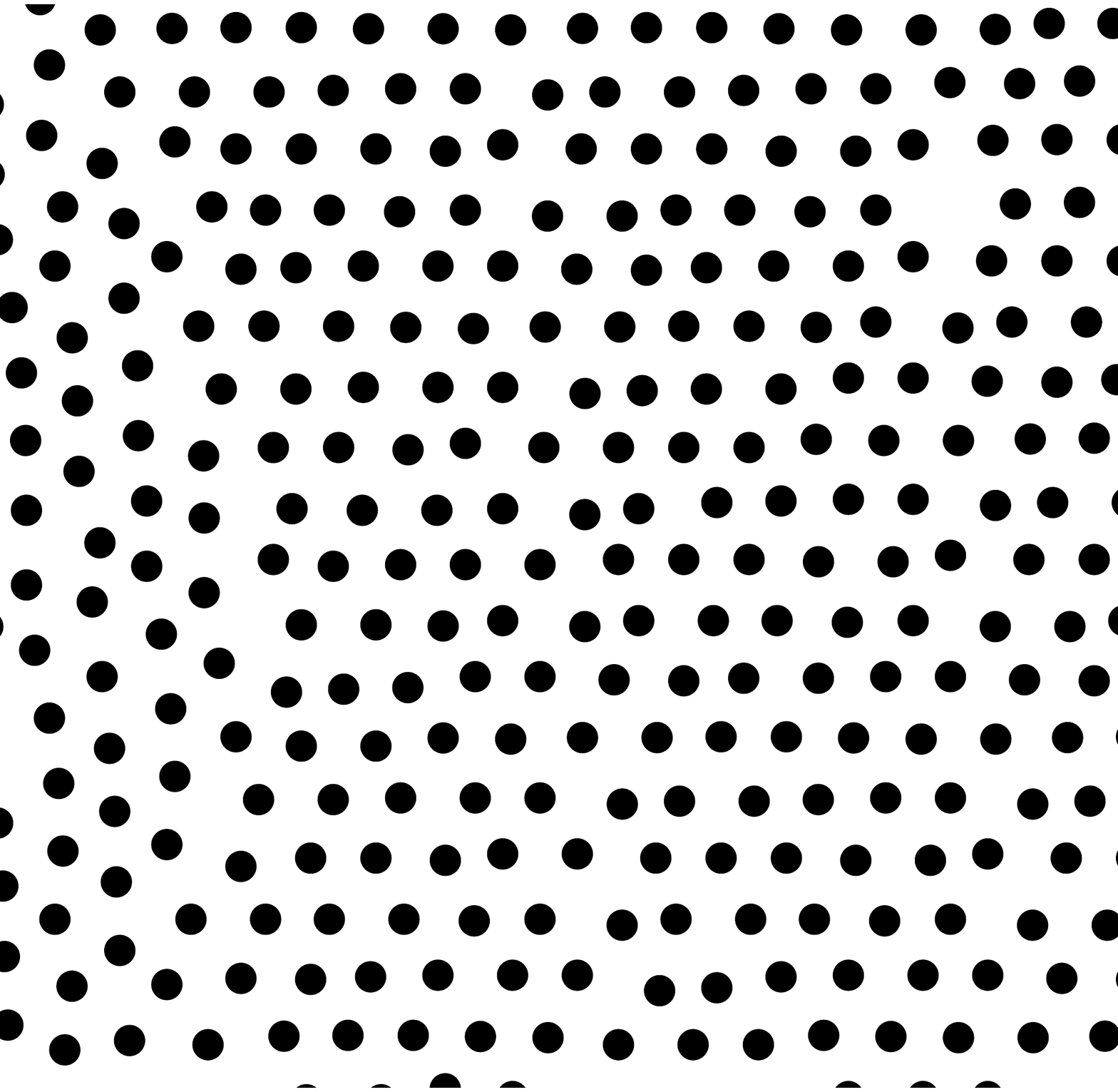}}&
\frame{\includegraphics[width=0.32\columnwidth,trim={0.cm 3.55cm 0.cm 3.55cm},clip]{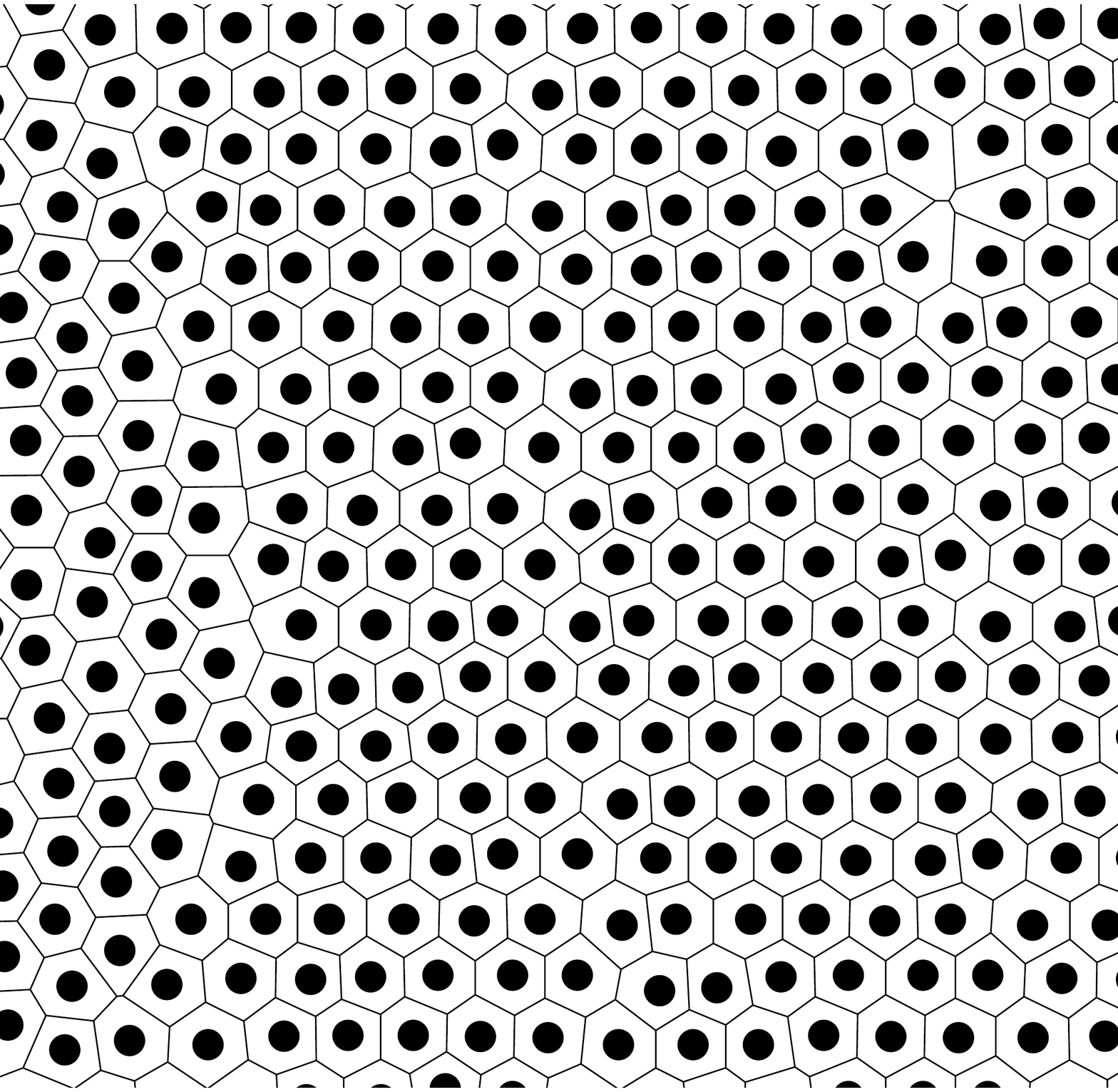}}&
\frame{\includegraphics[width=0.32\columnwidth,trim={0.cm 3.55cm 0.cm 3.55cm},clip]{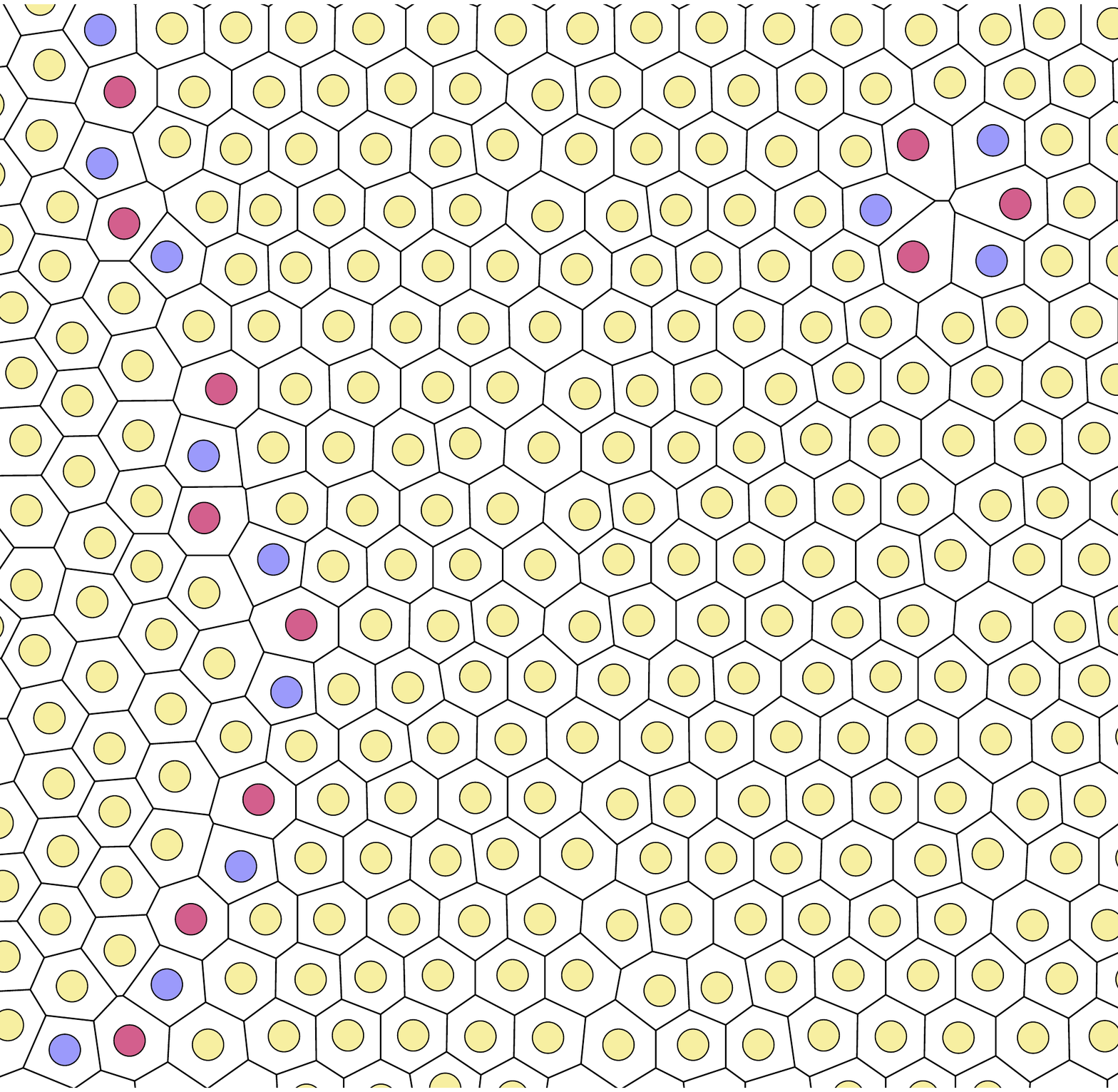}}\\
(a)&(b)&(c)
\end{tabular}
\caption{(a) A two-dimensional polycrystal; (b) atoms and their Voronoi cells; (c) atoms colored according to Voronoi topology.\label{all3}}
\end{center}
\end{figure}

Although {\it VoroTop} is designed to analyze three-dimensional systems, the theory behind it can be most clearly explained through consideration of two-dimensional ones.  Figure \ref{all3}(a) illustrates part of a two-dimensional polycrystalline system in which each point represents the position of an atom \cite{lazar2005u}.  A grain boundary on the left and a vacancy at the top right can be identified.  To make the rough notions of `crystalline' and `defect' precise, we consider the Voronoi tessellation of this system.

For a set of points in space, the {\it Voronoi cell} of each point is the region of space closer to that point than to any other \cite{voronoi1908nouvelles, okabe2009spatial}; this definition is equally applicable in two and three dimensions.  Figure \ref{all3}(b) shows the same system shown in Figure \ref{all3}(a) but with the Voronoi cell of each atom drawn.  Although many atoms near defects have five or seven edges, Voronoi cells of most atoms have six edges.  Indeed, the Voronoi cell of every atom in a defect-free hexagonal-lattice crystal has six edges, even at finite temperatures \cite{leipold2015statistical}.  This suggests defining crystalline atoms as those whose Voronoi cells have six edges, and defect atoms as those with more or fewer edges.  

A pair of atoms with adjacent Voronoi cells are defined as {\it neighbors}, and so the number of edges of a Voronoi cell can be understood as a count of neighbors.  Although this number often coincides with the ``coordination number'' (i.e., the number of atoms at most a given distance away), this is not always the case.  In two dimensions, we use the term topology to refer to the number of edges of a Voronoi cell.  This number describes the manner in which an atom's neighbors are arranged relative to a central atom and to one another.  

Coloring each atom according to its Voronoi cell topology highlights the defect structure in the polycrystal; cf. Figure \ref{all3}(c).  After identifying defects at the single-atom level, larger-scale defects can be defined as contiguous sets of individual defect atoms.  A ring of three red and three blue atoms, for example, identifies the presence of a vacancy.  Similarly, chains of alternating red and blue atoms indicate the presence of grain boundaries.

\subsection{Instability analysis and families of Voronoi topologies}
\label{ianalysis}

\begin{figure}
\begin{center}
\begin{tabular}[c]{ccc}
\frame{\includegraphics[width=0.2\columnwidth,trim={0.cm 3.55cm 0.cm 3.55cm},clip]{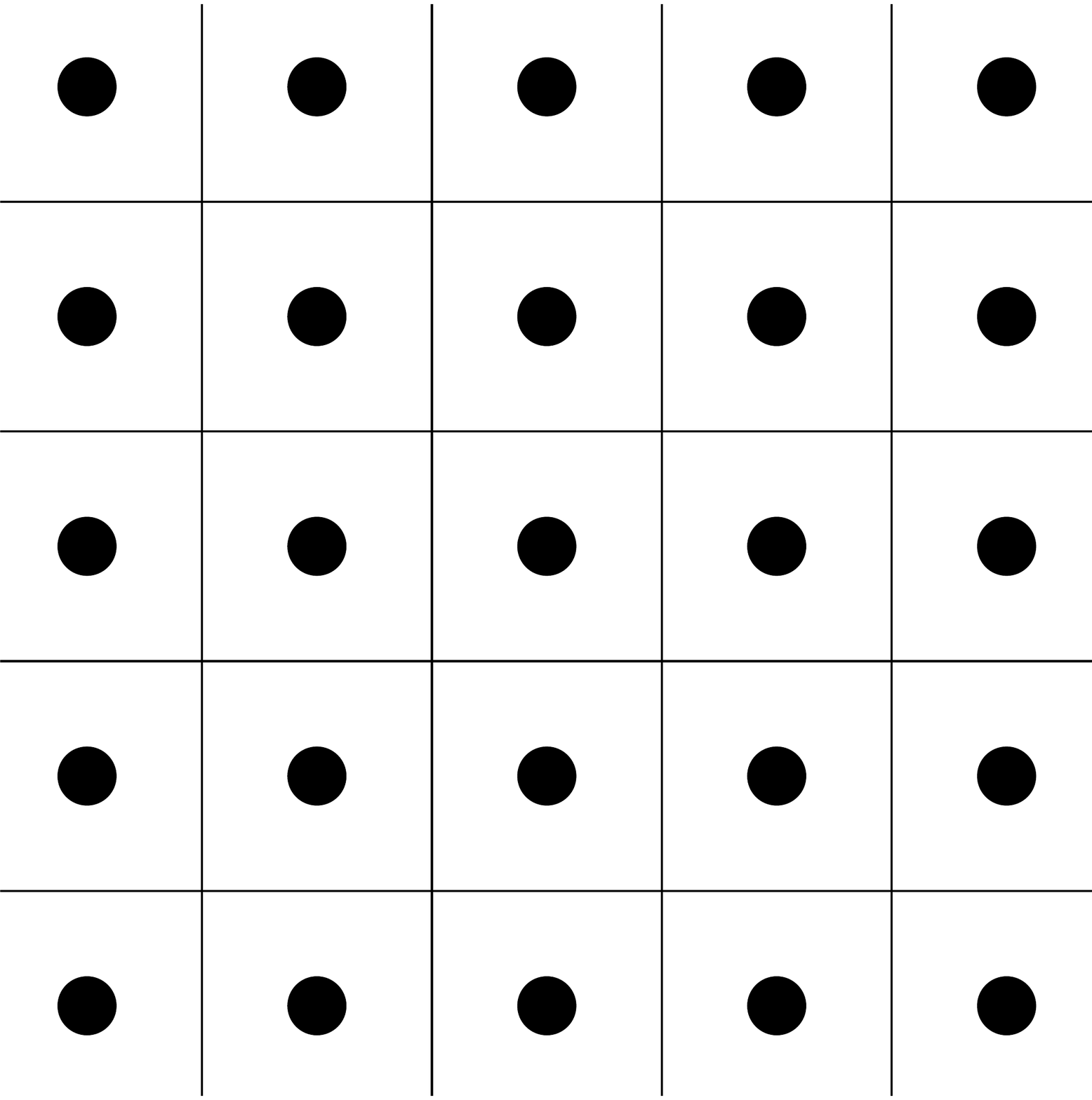}}&
\frame{\includegraphics[width=0.2\columnwidth,trim={0.cm 3.55cm 0.cm 3.55cm},clip]{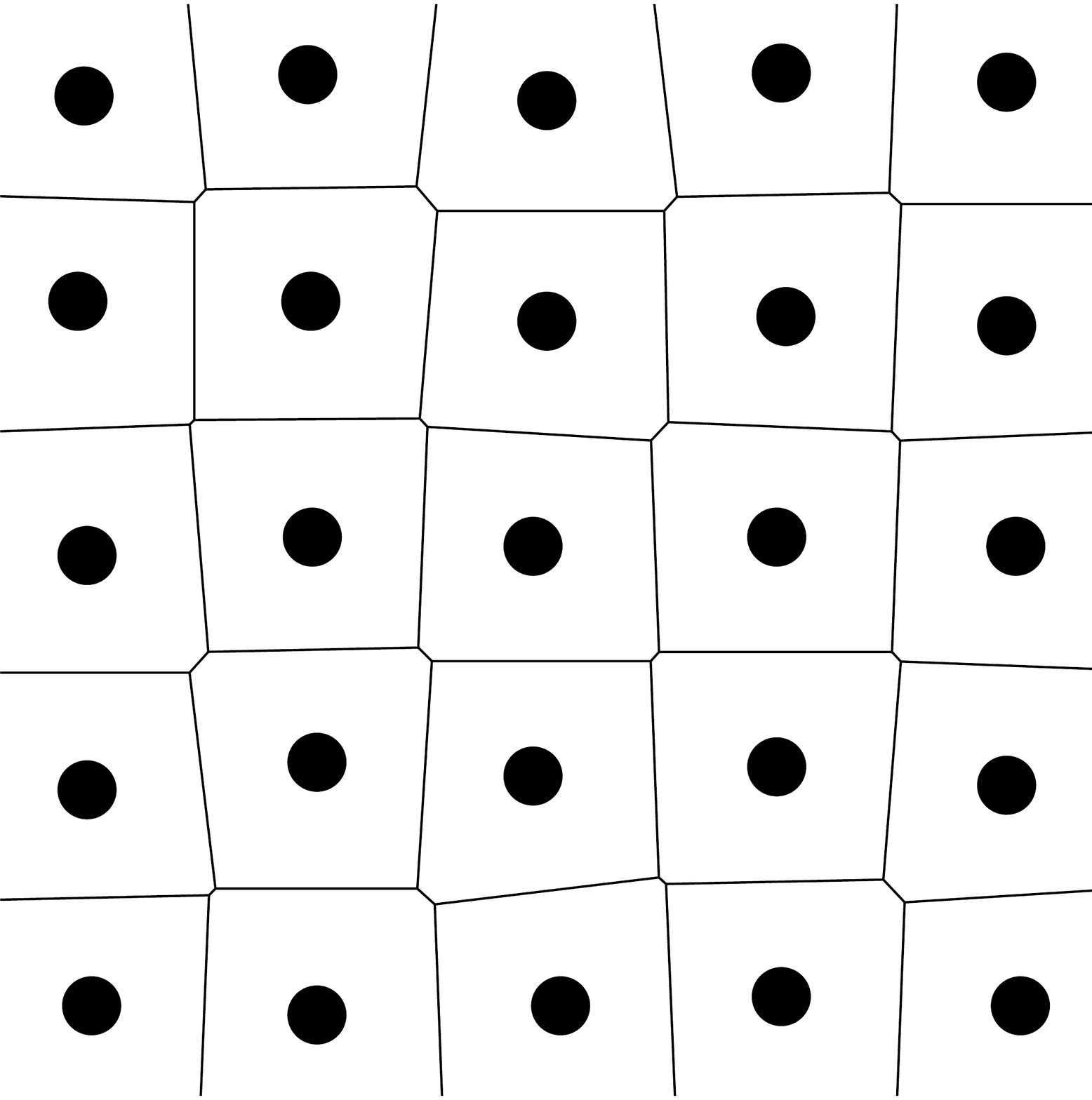}}&
\frame{\includegraphics[width=0.2\columnwidth,trim={0.cm 3.55cm 0.cm 3.55cm},clip]{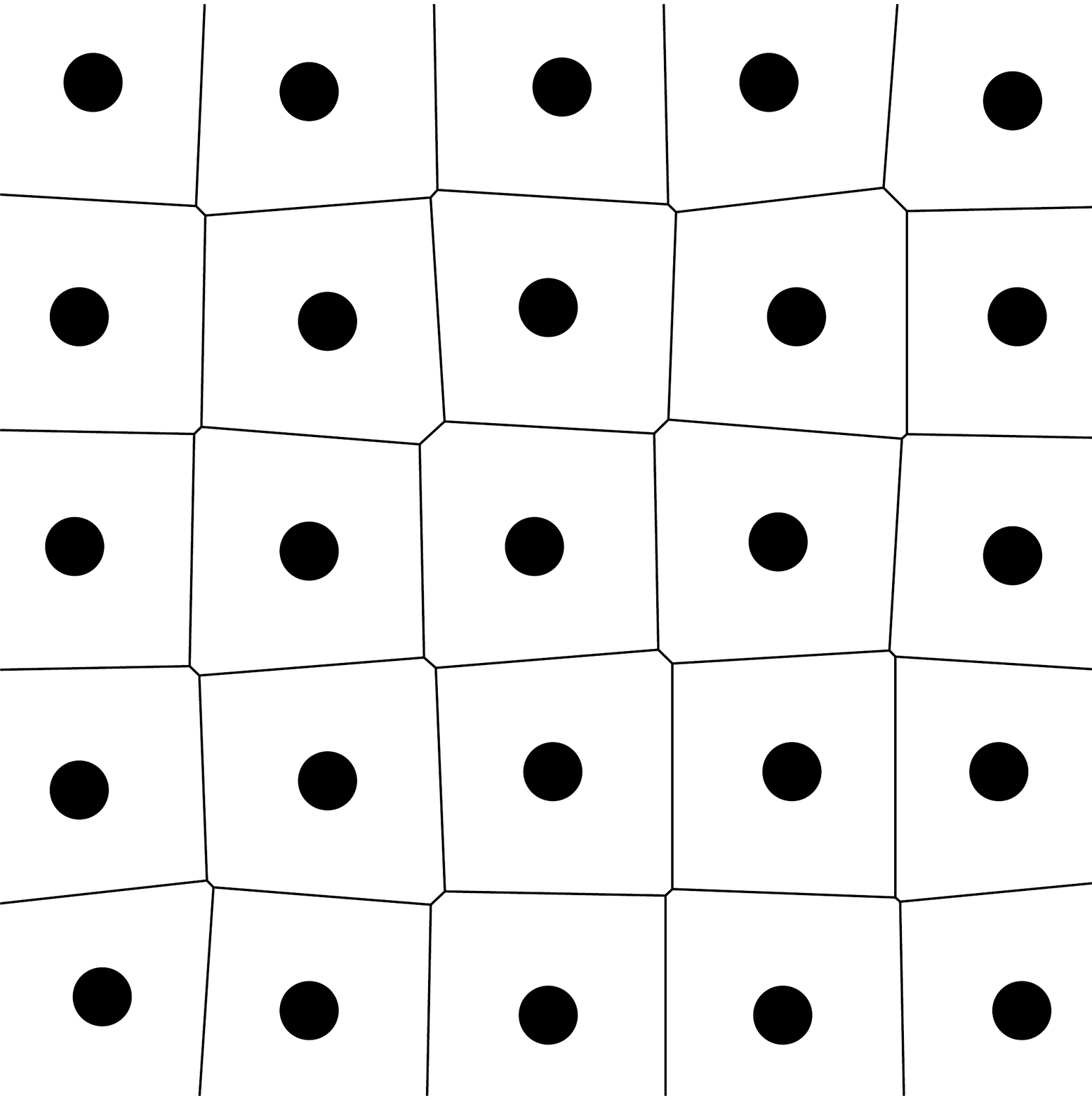}}\\
(a)&(b)&(c)
\end{tabular}
\caption{(a) Zero- and (b-c) finite-temperature square-lattice crystals. \label{squares}}
\end{center}
\end{figure}

Although small perturbations of an hexagonal lattice do not change the topology of its Voronoi cells, this is not the case for all lattices.  Consider, for example, the Voronoi cells in a two-dimensional square lattice, illustrated in Figure \ref{squares}(a). In a perfect crystal, the Voronoi cell of each atom has exactly four edges, and exactly four Voronoi cells meet at every corner.  These corners are unstable in the sense that small perturbations of the atomic coordinates change the topologies of nearby cells; cf.~Figures \ref{squares}(b) and (c).  A consequence of this instability is that Voronoi cells in finite-temperature square-lattice crystals can have between 4 and 8 edges.  We use the term {\it family} to refer to a set of topologies that can be obtained from an ideal structure through infinitesimal perturbations of atomic coordinates.  Atoms whose Voronoi topologies belong to such a set are classified as belonging to a crystal, and atoms with other topologies are classified as belonging to defects.  

\subsection{Three dimensions}
\label{three}

\begin{figure}[b]
\centering
\includegraphics[width=0.3\columnwidth]{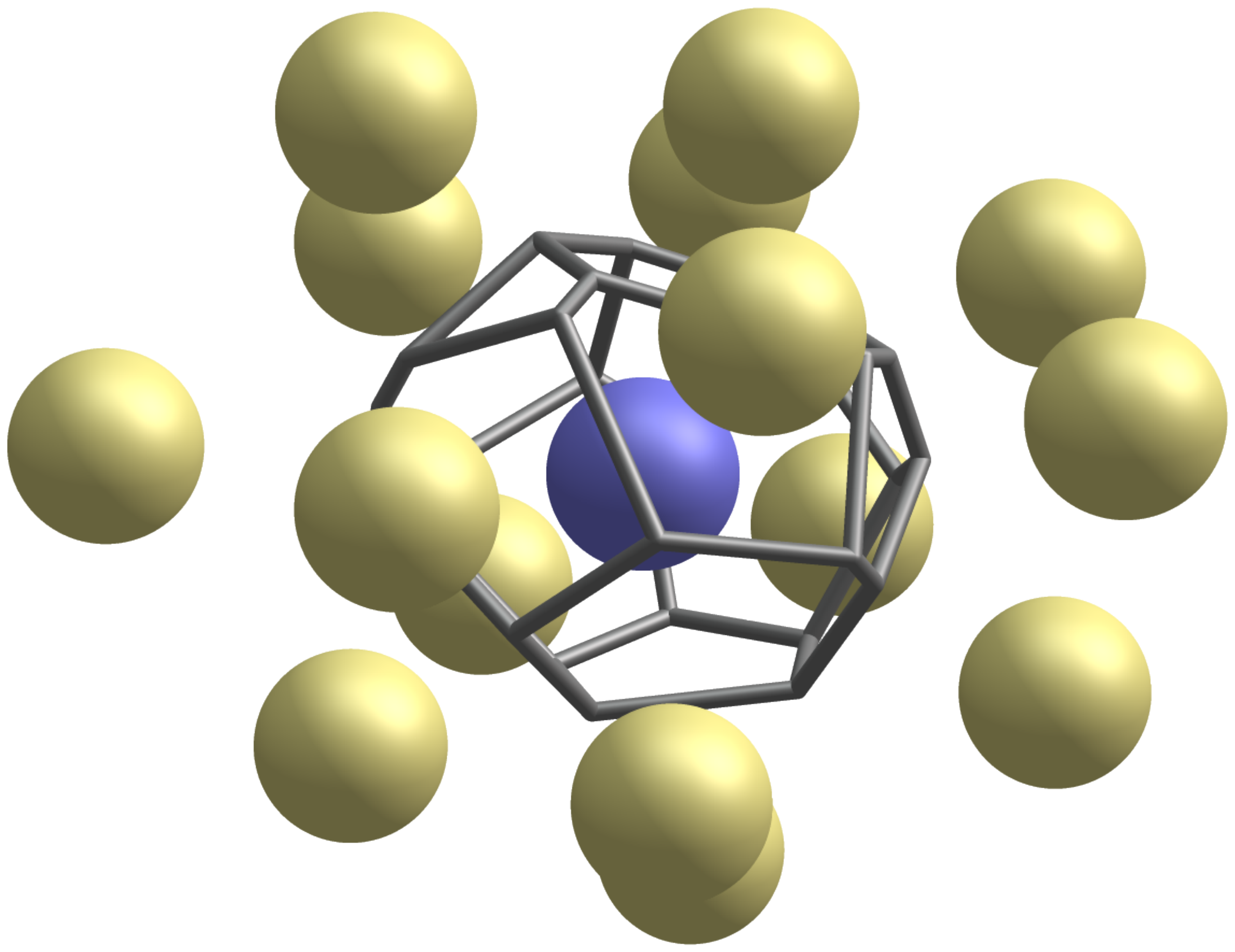}
\caption{The Voronoi cell of a central atom surrounded by neighboring atoms.}
\label{voronoi}
\end{figure}

The analysis described in Sections \ref{vctls} and \ref{ianalysis} can be applied directly to the study of three-dimensional systems.  Figure \ref{voronoi} illustrates a central blue atom, its Voronoi cell, and neighboring gold atoms.  In three dimensions, we use the term {\it Voronoi topology} to refer not only to the number of faces of a Voronoi cell, but also to the manner in which those faces are arranged.  
This information describes the manner in which an atom's neighbors are arranged relative to a central atom and to one another.  For example, the number of sides of a face indicates the number of neighbors shared in common between a central atom and one of its neighbors.  Details about how the Voronoi topology can be efficiently computed and recorded are reported elsewhere \cite{2012lazar, lazar2011evolution}.

We use Voronoi topology to define crystal and defect structure in three dimensions in the same manner as we did in two dimensions.  In particular, to a given ideal structure we associate a family of Voronoi topologies that can be obtained from infinitesimal perturbations of atomic coordinates.  Then, when analyzing data, we compute the topology of each Voronoi cell.  Atoms whose Voronoi cell topologies belong to the given family are considered crystalline; atoms with other topologies are considered defects.  

One complication that arises in this approach is the possibility that one Voronoi topology belongs to multiple families.  For example, the two-dimensional Voronoi cell topology with 6 edges belongs to both the square- and hexagonal-lattice families.  This indeterminacy complicates the analysis of certain systems.  A method of resolving this indeterminacy is described in Section \ref{indeterminate}.

\section{Extensions}
\label{extensions}

\subsection{Determining families of topologies}
\label{filterdev}

A family of Voronoi topologies associated with a given structure can be determined analytically by considering the changes that can occur near each corner of an unstable Voronoi cell \cite{lazar2015topological}.  For example, in three-dimensional face-centered cubic (FCC) crystals, each of six unstable corners can resolve in one of eight ways, meaning that up to $8^6 =$ 262,144 topologies must be considered.  However, many of these resolutions are equivalent, and ultimately there are only 6294 unique topologies \cite{lazar2015topological}.  In hexagonal close-packed (HCP) crystals, similar analysis shows that there are 21,611 unique topologies.  

Although the analytic determination of a family of Voronoi topologies is possible in many cases, it can sometimes be computationally prohibitive.  For example, each corner of a Voronoi cell in the three-dimensional simple cubic lattice can resolve in one of 32 distinct ways.  Since each cell has 8 unstable corners, $32^8 \approx 10^{12}$ possible resolutions must be considered.  Aside from the significant computing power necessary to initially determine this family, storing it in memory would also be prohibitive.  We therefore consider a Monte Carlo approach to determining families.  In particular, we take an ideal system and apply a small random perturbation to each atomic coordinate -- mimicking the effect of finite temperature -- and record the set of Voronoi topologies in the perturbed sample.  As we consider increasingly many samples, we more accurately approximate the family of topologies.  In addition to computational tractability, another advantage of this numerical approach over the analytic one is its applicability to more general structures than lattices.  For example, it can be used to study the complex structure of grain boundaries.

\subsection{Resolving indeterminate Voronoi topologies}
\label{indeterminate}

At the end of Section \ref{three} we noted the possibility that a single Voronoi topology will belong to multiple families; we call such Voronoi topologies {\it indeterminate}.  As a concrete example, many Voronoi topologies that belong to the FCC family also belong to the HCP family.  This complicates the analysis of FCC systems which contain defects such as stacking faults, since the local structure of atoms associated with those defects resembles that of HCP crystals.  We first illustrate the extent of the problem, and then explain how we handle it.

\begin{figure}
\setlength{\tabcolsep}{2.2pt}
\begin{center}
\begin{tabular}{ccc}
\frame{\includegraphics[width=0.3\columnwidth]{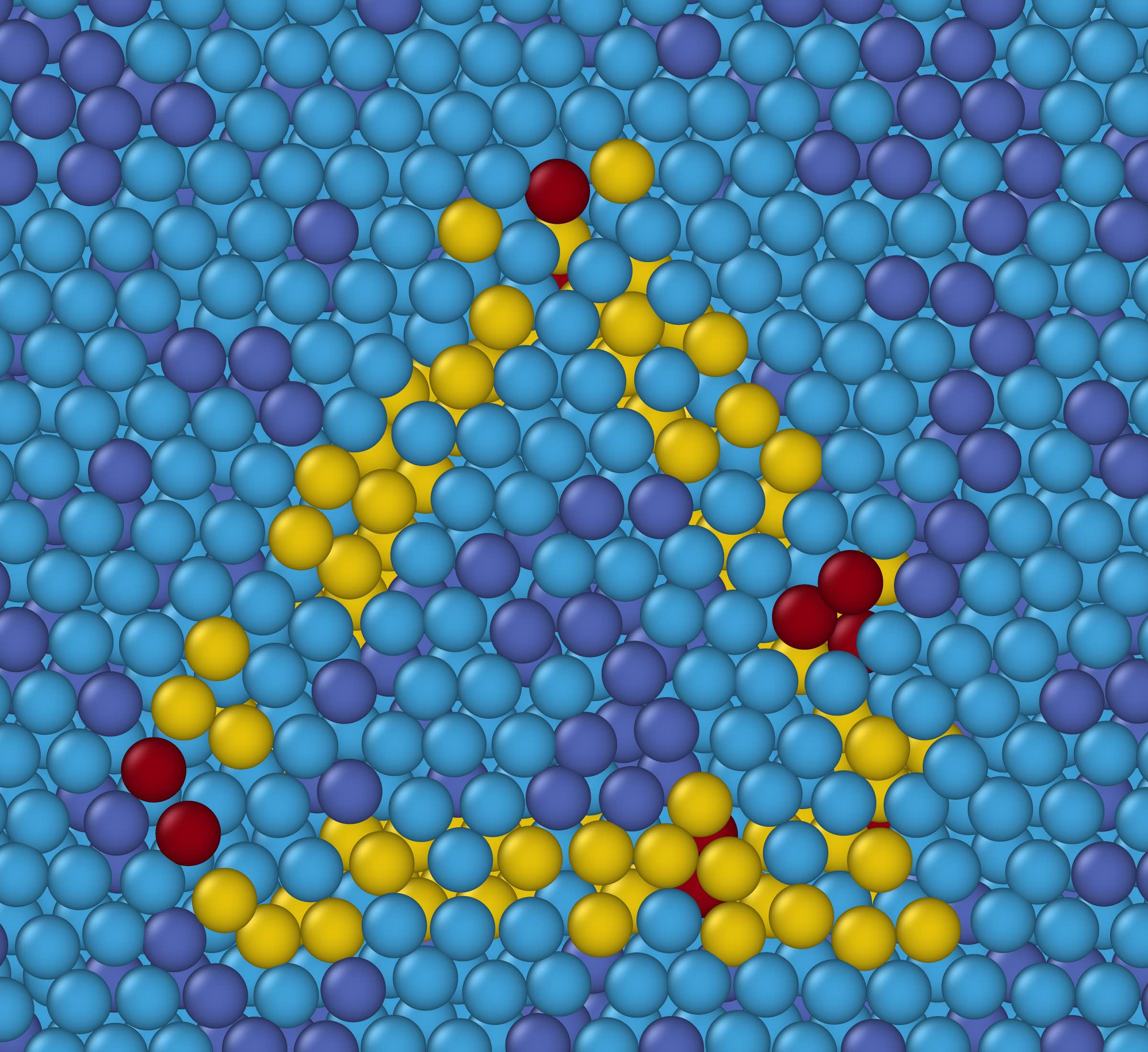}}&
\frame{\includegraphics[width=0.3\columnwidth]{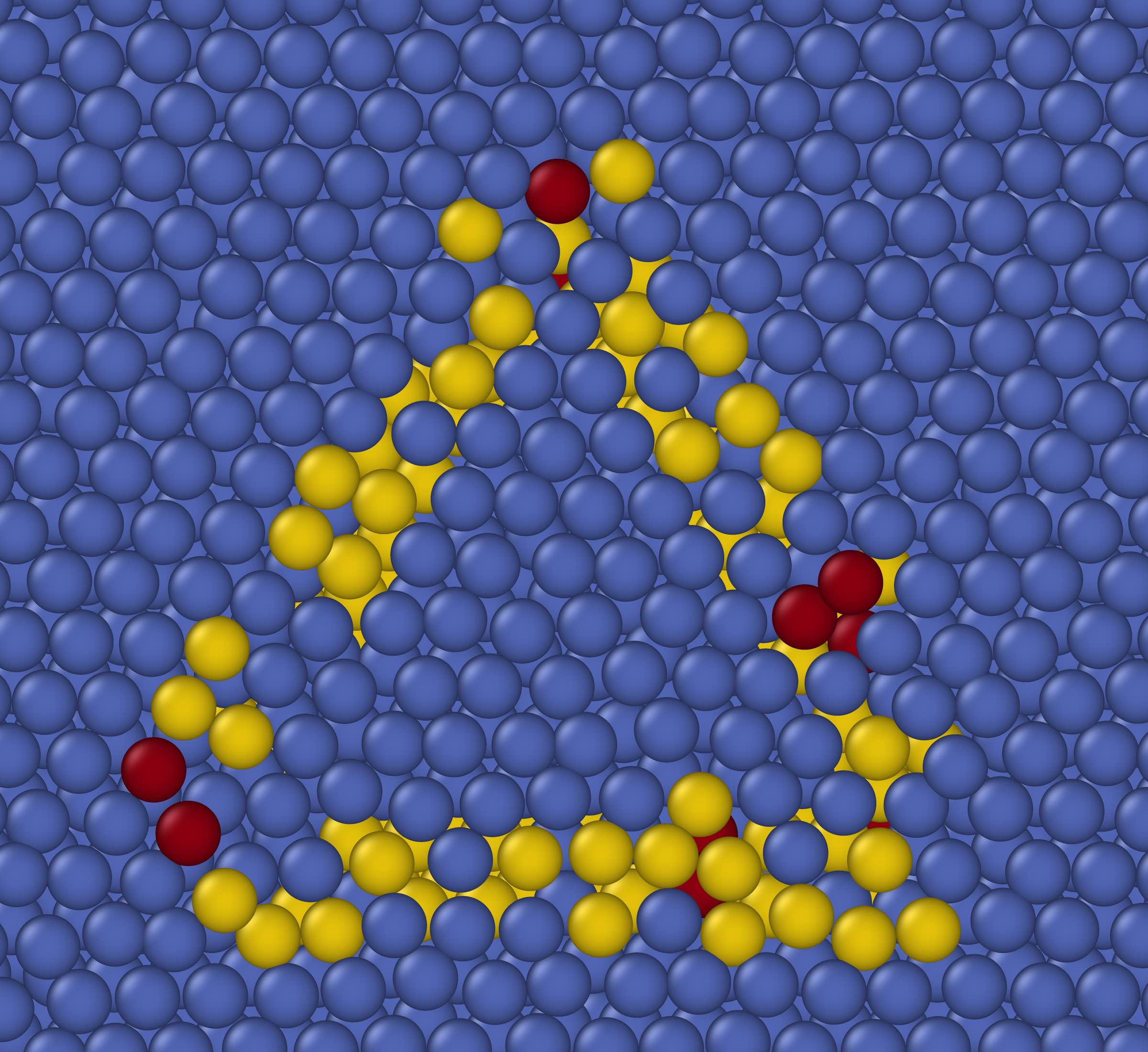}}&
\frame{\includegraphics[width=0.3\columnwidth]{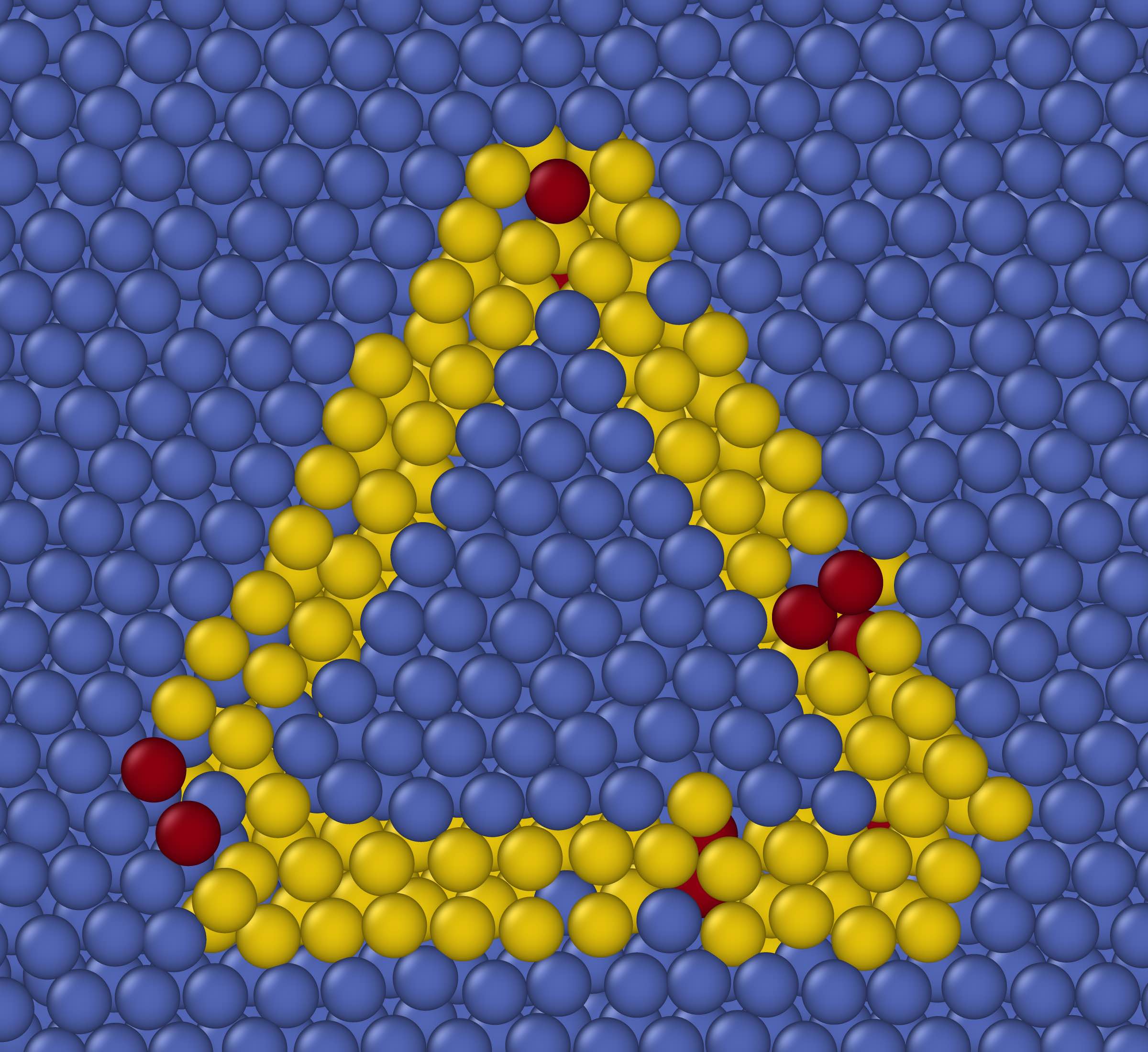}}\\
(a)&(b)&(c)
\end{tabular}
\caption{A stacking-fault tetrahedron; dark blue atoms indicate FCC local structure; yellow atoms indicate HCP local structure; red atoms indicate other topologies; light blue indicates FCC-HCP indeterminate topologies. In (b), atoms with indeterminate Voronoi topologies are treated as having FCC local structure.  In (c), indeterminate topologies are resolved using a perturbation analysis described in the text. \label{indet}}
\end{center}
\end{figure}

Figure \ref{indet} shows a cross-section of a stacking-fault tetrahedron in a copper crystal heated to 85\% of its bulk melting temperature.  In Figure \ref{indet}(a), atoms with determinate FCC local structure are colored dark blue, those with determinate HCP local structure are colored yellow, and those with indeterminate FCC-HCP topologies are colored light blue; all remaining atoms are colored red.  Indeterminate topologies clearly account for a large number of atoms.  

One way to approach this problem is to treat all atoms with indeterminate topologies as having FCC local structure, and to identify only atoms with determinately HCP topologies as having HCP local structure.  Figure \ref{indet}(b) shows all indeterminate topologies colored as FCC atoms.  Although this approach mischaracterizes many HCP atoms as having FCC local structure, it still produces a relatively clear picture of the defect.  

Another way to approach this problem is computationally more demanding but ultimately more accurate.  Instead of treating all indeterminate topologies as having FCC local structure, we consider how the topology of a Voronoi cell changes when atomic coordinates are perturbed.  If the resulting topologies of an atom tend to be determinately FCC, then we identify that atom as having FCC local structure; if they tend to be determinately HCP then we identify it as having HCP local structure.  The effectiveness of this approach is best understood in the context of the underlying configuration-space theory \cite{lazar2015topological}.

Figure \ref{indet}(c) shows results of this analysis.  The coordinates of each atom were displaced by a random perturbation chosen from a normal distribution 0.05 times the cubed root of the volume of its Voronoi cell.  Voronoi topologies of atoms with initially indeterminate topologies were computed in this perturbed system, and the process was repeated five times.  If more perturbations resulted in determinate HCP topologies of a particular atom than in determinate FCC topologies, then that atom was identified as having HCP local structure; otherwise, the atom was identified as having FCC local structure.  This analysis provides a clear picture of the stacking-fault tetrahedron.

\subsection{Cluster analysis} 
\label{clusteranalysis}

The analysis explained in Section \ref{basics} results in a classification of atoms as either belonging to a crystal or to a defect.  This approach by itself, however, does not say anything about defects consisting of multiple atoms.  Even a vacancy or interstitial, which affect the ordering near several neighboring atoms, are not explicitly defined.  One approach towards precisely defining large-scale defects involves the consideration of clusters of defect atoms.  

\begin{figure}
\setlength{\tabcolsep}{6pt}
\begin{center}
\begin{tabular}{ccc}
\frame{\includegraphics[width=0.2\columnwidth]{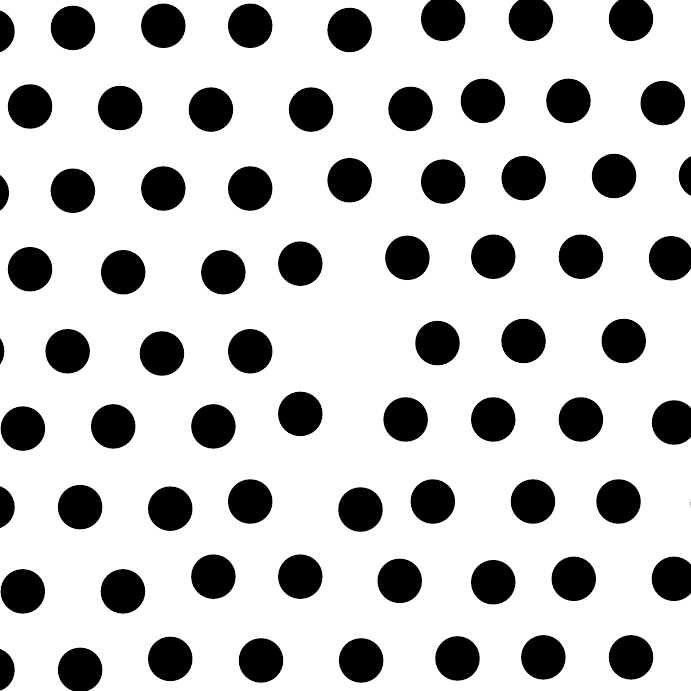}}&
\frame{\includegraphics[width=0.2\columnwidth]{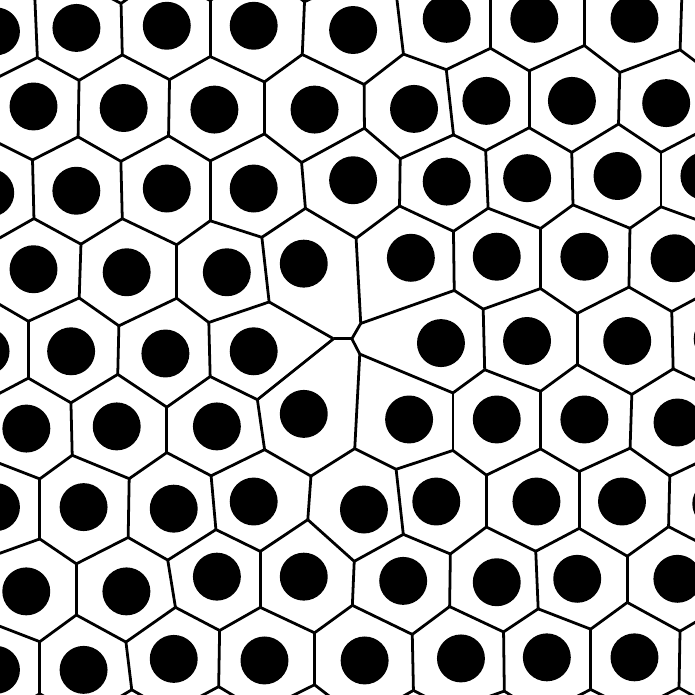}}&
\frame{\includegraphics[width=0.2\columnwidth]{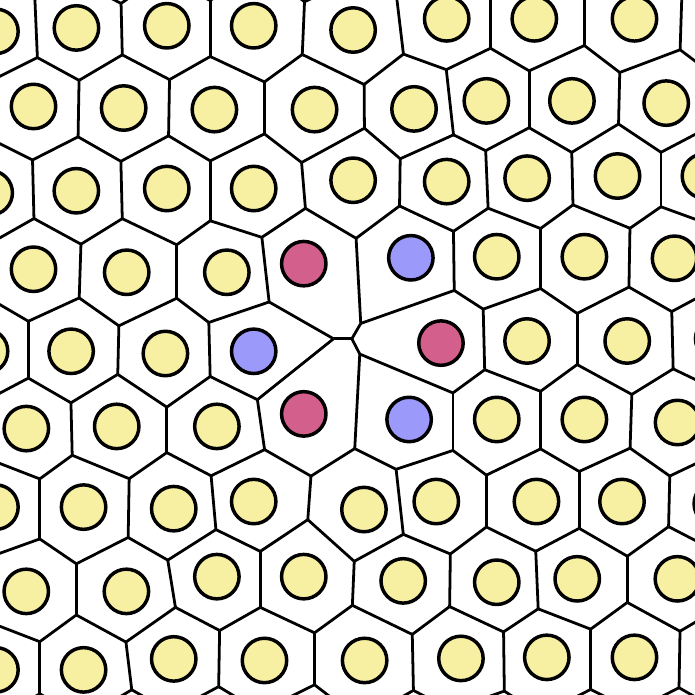}}
\end{tabular}
\caption{A cluster of defect atoms associated with a vacancy.\label{vacancy}}
\end{center}
\end{figure}

We define a defect {\it cluster} to be a set of neighboring defect atoms.  To illustrate this idea, Figure \ref{vacancy} shows a vacancy seen earlier in Figure \ref{all3}.  The six adjacent defect atoms are defined as belonging to a single defect cluster associated with the vacancy.  Further analysis of the topologies that constitute a cluster might provide a more informative representation of larger-scale defect structure.  For example, it is no coincidence that the number of edges of the Voronoi cells in this defect cluster sum to a multiple of 6.  This is a mathematical result about simple planar graphs, and consequently about any defect cluster located in an otherwise perfect finite-temperature hexagonal-lattice crystal.

\section{Comparison with other methods}
\label{compare}

\begin{figure}
\setlength{\tabcolsep}{2.2pt}
\begin{center}
\begin{tabular}{cc}
\frame{\includegraphics[width=0.3\columnwidth]{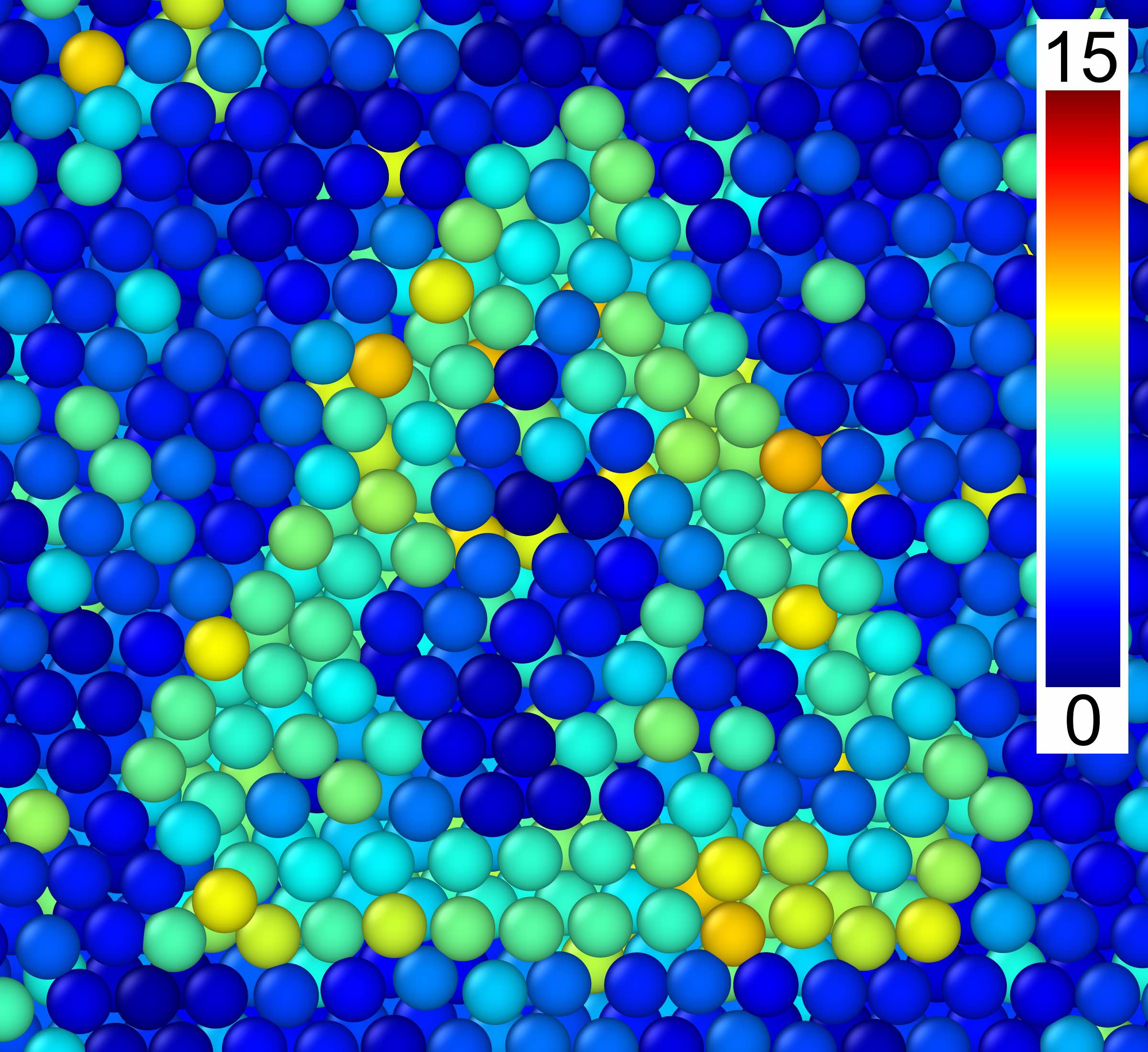}}&
\frame{\includegraphics[width=0.3\columnwidth]{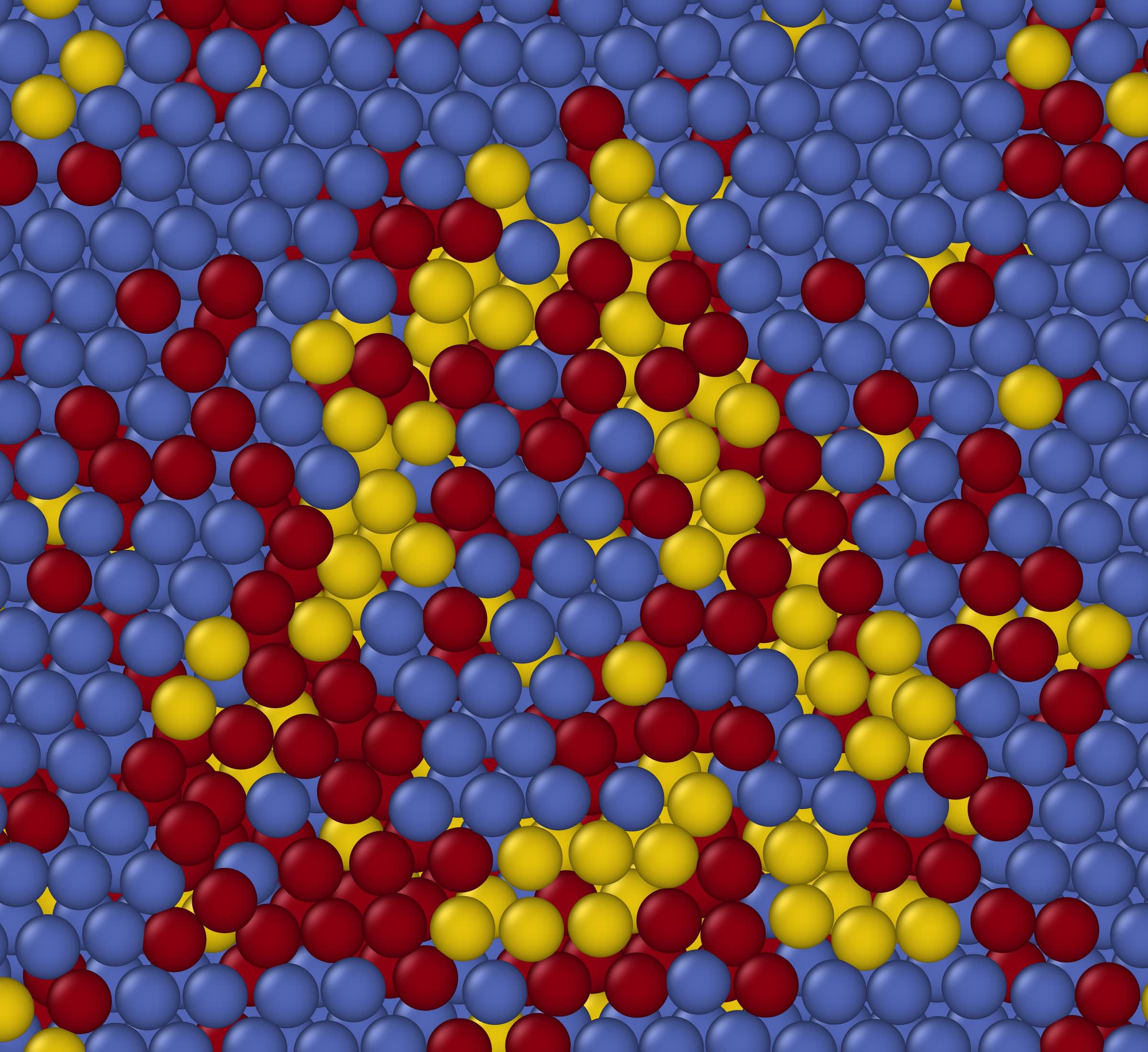}}\\
(a) Centrosymmetry &(b) Bond-angle analysis \vspace{2mm}\\
\frame{\includegraphics[width=0.3\columnwidth]{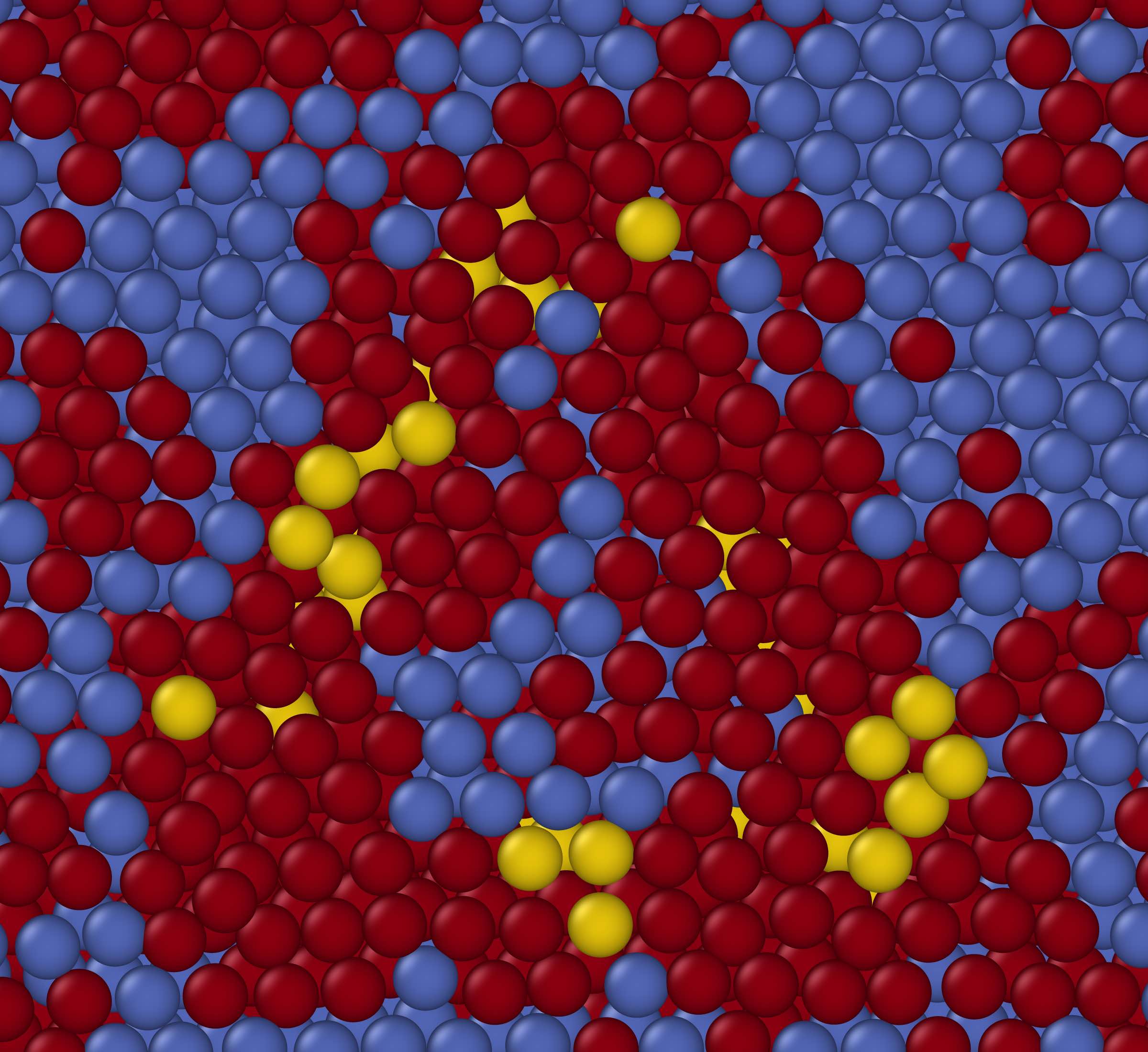}}&
\frame{\includegraphics[width=0.3\columnwidth]{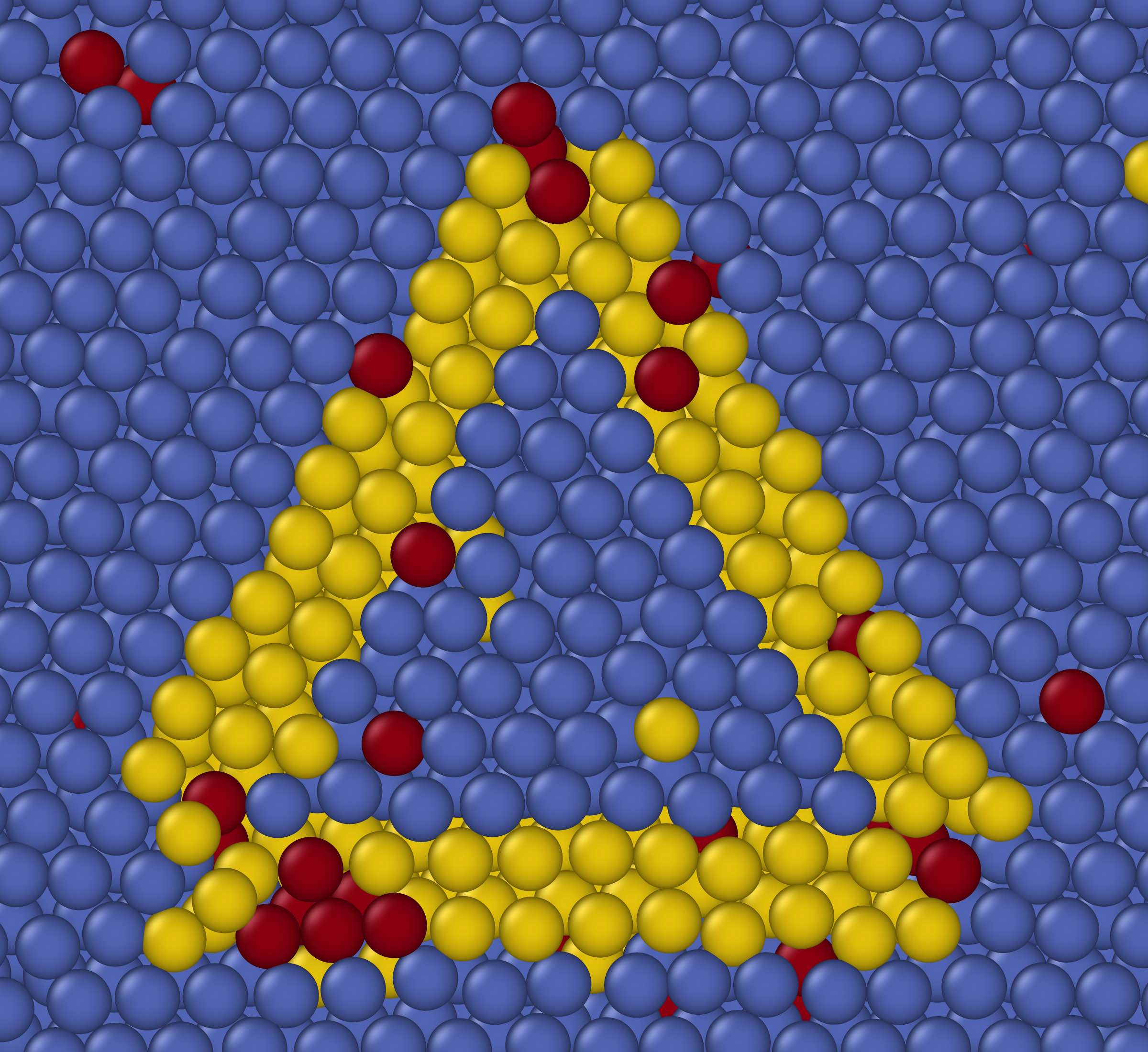}}\\
(c) Adaptive CNA &(d) PTM
\end{tabular}
\caption{A stacking-fault tetrahedron visualized using (a) centrosymmetry, (b) bond-angle analysis, (c) adaptive common-neighbor analysis, and (d) polyhedral template matching.  In (b), (c), and (d) blue atoms indicate FCC local structure, yellow atoms indicate HCP local structure, and red atoms indicate other local structure.  \label{sft}}
\end{center}
\end{figure}

We briefly compare Voronoi topology with three conventional approaches to local structure analysis, and a very recent one.  Figure \ref{sft} illustrates the same stacking-fault tetrahedron considered in Section \ref{indeterminate} visualized using centrosymmetry \cite{kelchner1998dislocation}, bond-angle analysis \cite{ackland2006applications}, adaptive common-neighbor analysis \cite{stukowski2012structure, honeycutt1987molecular}, and polyhedral template matching \cite{larsen2016robust}.  While the general shape of the stacking-fault tetrahedron can be discerned in (a), (b), and (c), its details are ambiguous.  In particular, many atoms belonging to the defect are misidentified as belonging to the bulk, and many atoms in the bulk are misidentified as having non-crystalline structure.  These results are in sharp contrast with the picture produced using Voronoi topology, illustrated in Figure \ref{indet}.  

A new structure-analysis tool called polyhedral template matching (PTM), which also uses topology to describe local structure in atomic systems, has been recently introduced by Larsen et al.~\cite{larsen2016robust}.  PTM provides a picture of the stacking-fault tetrahedron that is comparable in quality to that produced by Voronoi topology; cf. Figure \ref{sft}(d).

\section{Case study: dislocation identification}
\label{case}

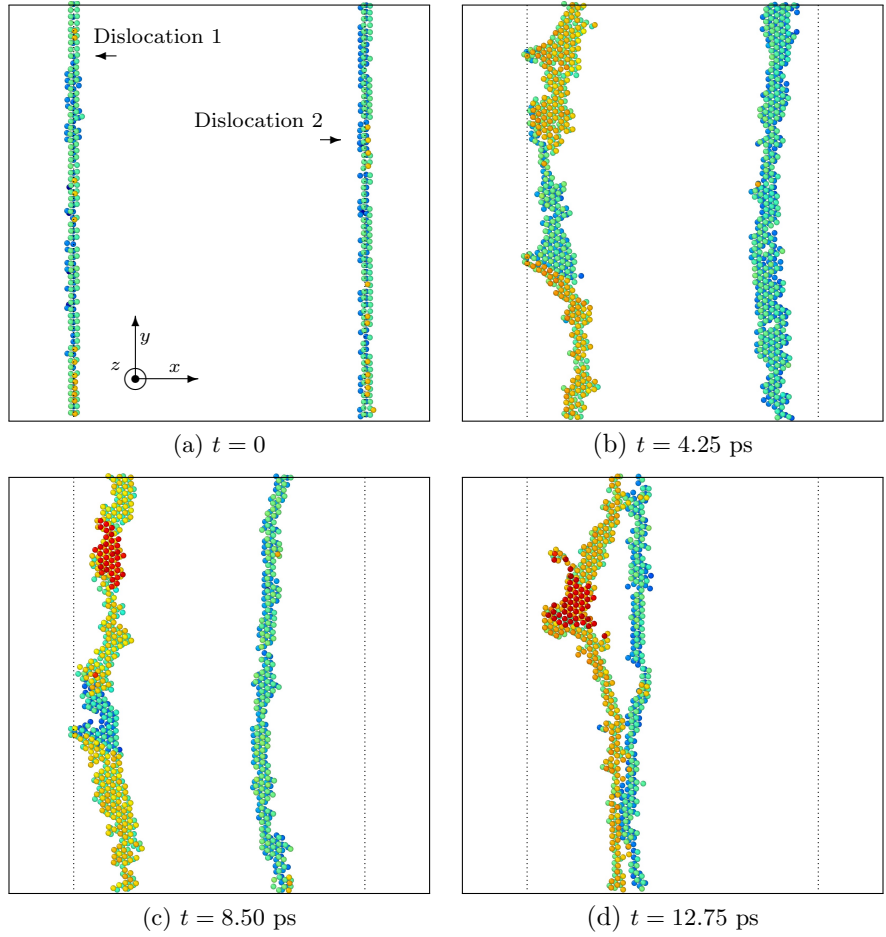
\begin{figure}
\begin{center}
\begin{tabular}[c]{cc}
\fbox{\begin{overpic}[width=0.46\columnwidth]{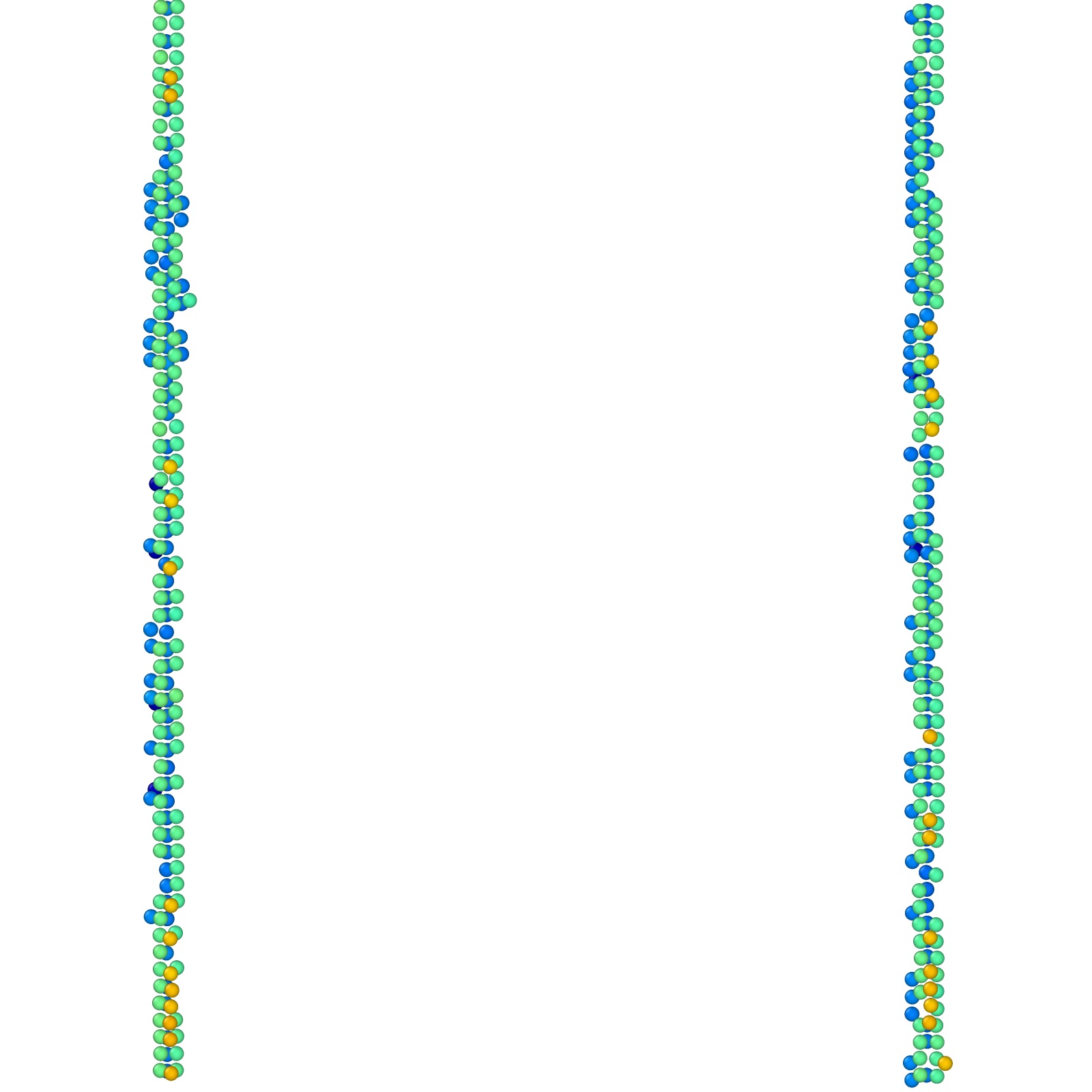}
\dottedline{1}(15.3,1)(15.3,99)
\dottedline{1}(84.7,1)(84.7,99)
\put(30,10){\vector(1,0){15}}
\put(30,10){\vector(0,1){15}}
\put(30,10){\circle{5}}
\put(30,10){\circle*{2}}
\put (38,11.5) {\scriptsize $x$}
\put (31,19) {\scriptsize $y$}
\put (24,12) {\scriptsize $z$}
\put (20.00,90) {\footnotesize Dislocation 1}
\put (25.50,87) {\vector(-1,0){5}}
\put (44.00,70) {\footnotesize Dislocation 2}
\put (74.00,67) {\vector(1,0){5}}
\end{overpic}} &
\fbox{\begin{overpic}[width=0.46\columnwidth]{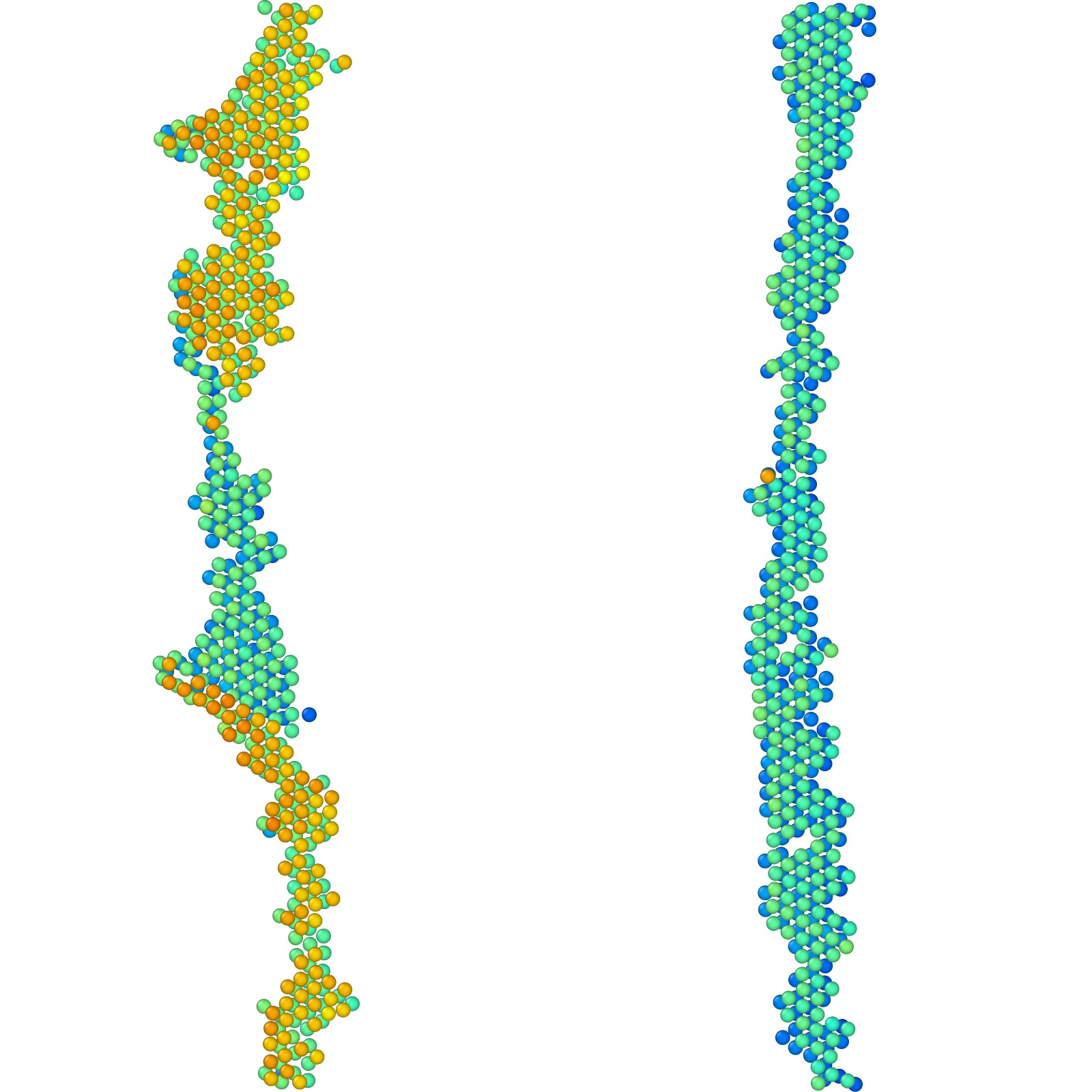}\dottedline{1}(15.3,1)(15.3,99)\dottedline{1}(84.7,1)(84.7,99)
\end{overpic}}\\
{\small (a) $t=0$} & (b) {\small $t=4.25$ ps}
  \\[2mm]
\fbox{\begin{overpic}[width=0.46\columnwidth]{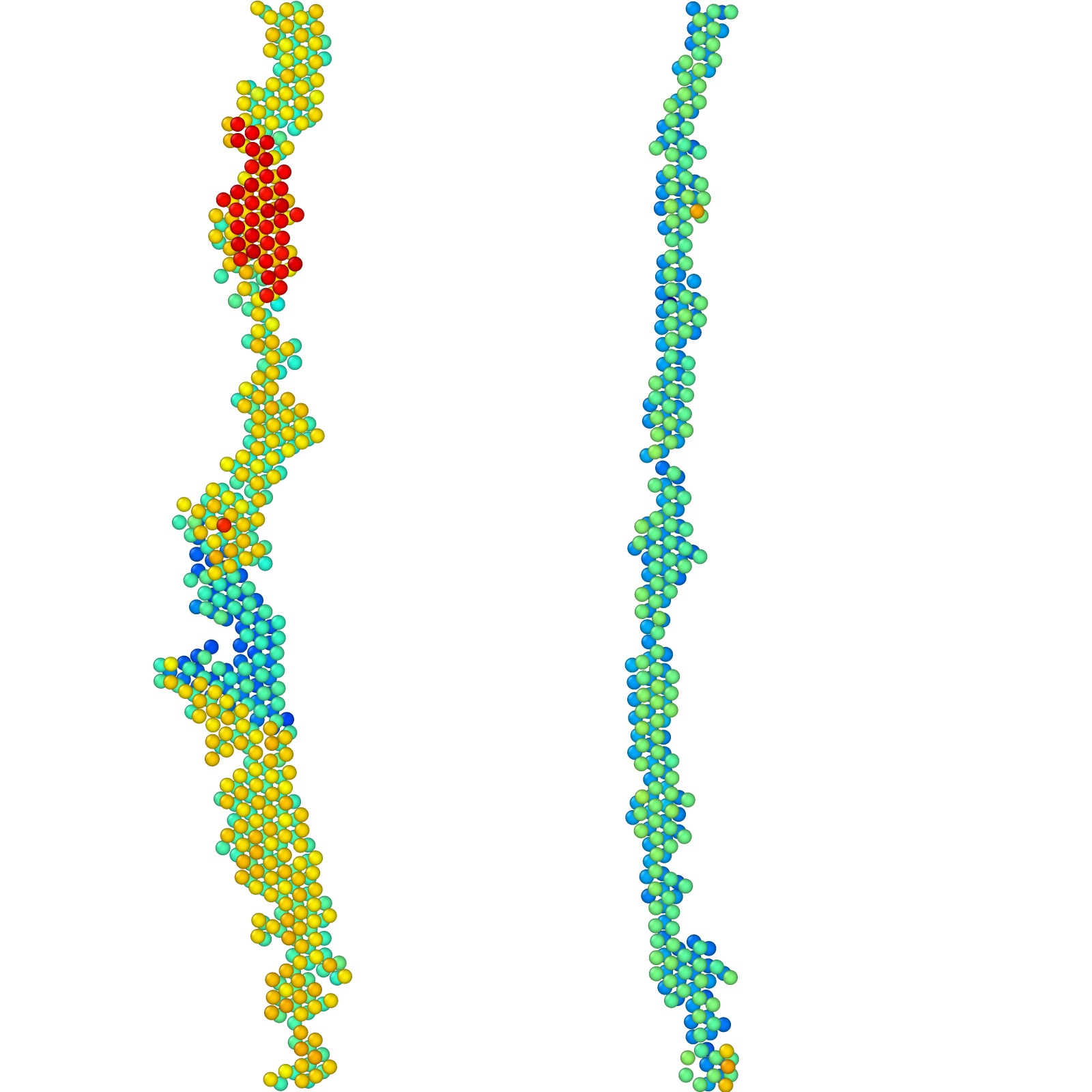}\dottedline{1}(15.3,1)(15.3,99)\dottedline{1}(84.7,1)(84.7,99)
\end{overpic}} &
\fbox{\begin{overpic}[width=0.46\columnwidth]{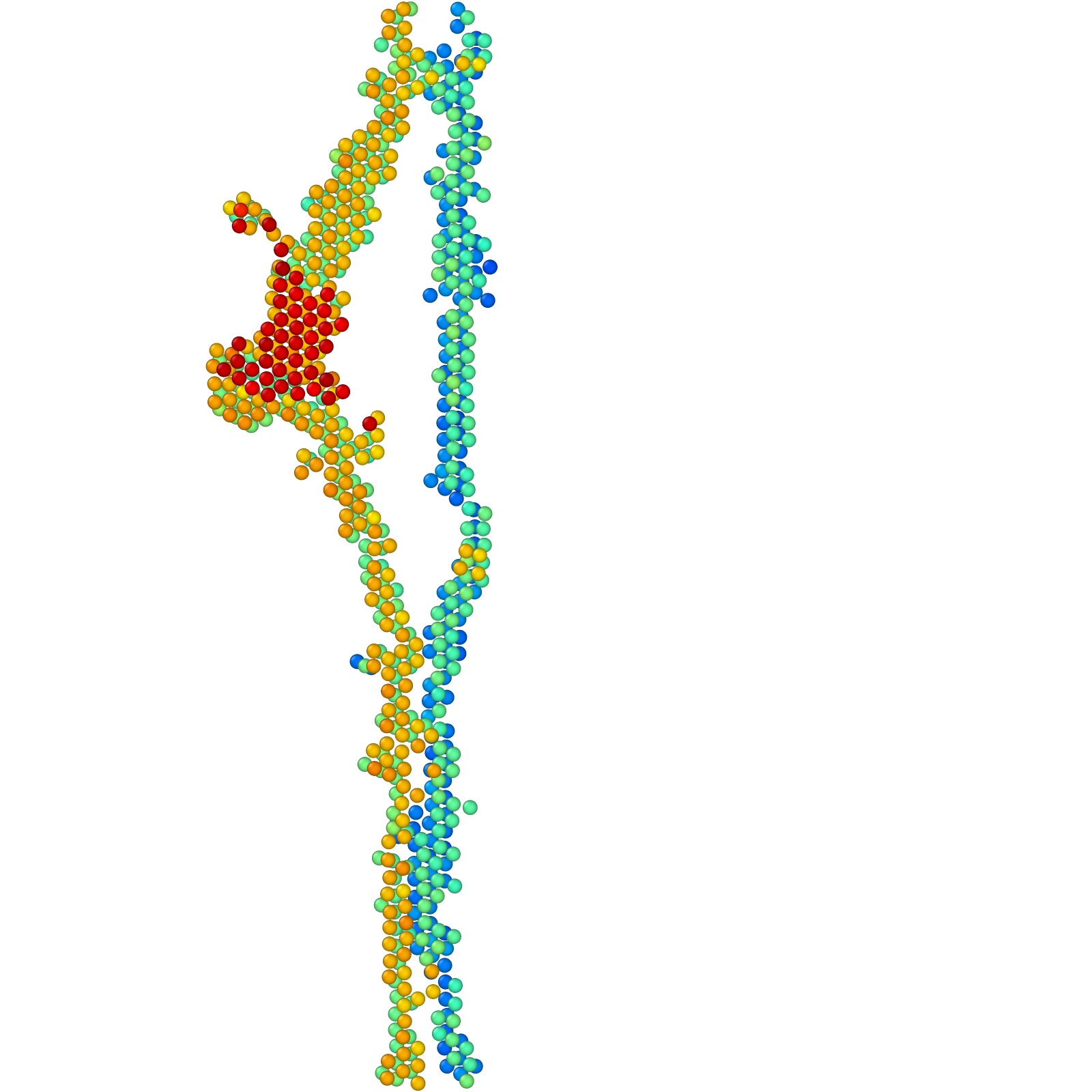}\dottedline{1}(15.3,1)(15.3,99)\dottedline{1}(84.7,1)(84.7,99)
\end{overpic}} \\
{\small (c) $t=8.50$ ps} & (d) {\small $t=12.75$ ps}
\end{tabular}
\end{center}
\caption{Two screw dislocations with Burgers vectors $\frac{a}{2}[1\bar{1}0]$ and $\frac{a}{2}[\bar{1}10]$ migrating towards each other in a $(111)$ slip plane.  Atoms are colored according to their coordinate along the $z$-axis; dotted lines indicate initial positions of dislocations.  \label{screws}}
\end{figure}

We use {\it VoroTop} to visualize dislocations at high-temperature and to analyze their behavior.  In particular, we consider a pair of screw dislocations with Burgers vectors $\frac{a}{2}[1\bar{1}0]$ and $\frac{a}{2}[\bar{1}10]$ migrating towards each other in an FCC aluminum crystal heated to 95\% of its melting temperature.  Figures \ref{screws}(a-d) illustrate a time-series of the evolution; a video showing the entire evolution can be found online in Supplementary Materials.  Atoms with Voronoi topologies belonging to the FCC family are not shown; all other atoms are colored according to their $z$ coordinate, to provide a clearer picture of their spatial arrangement. 
\setlength{\fboxsep}{0pt}

Although the dislocations initially lie in the same plane, portions of Dislocation 1 slip to other planes during the evolution, as can be seen in the changing colors of its constituent atoms.  These cross-slips decrease the effective mobility of this dislocation as compared with that of Dislocation 2, which remains in the same plane throughout the evolution.  The effect of cross-slips in pinning the dislocation mobility can be seen most clearly in the online video.

\begin{figure}
\begin{center}
\begin{overpic}[width=\columnwidth]{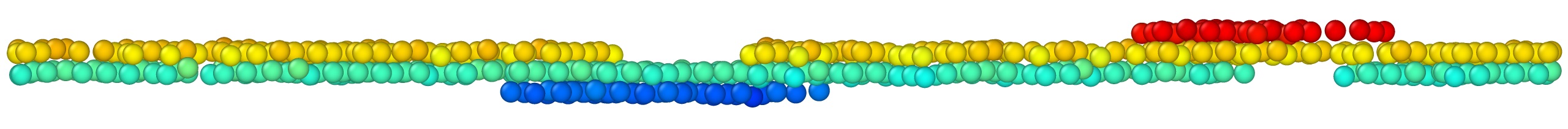}
\put (1,7) {\small (a)}
\put(93,8){\vector(1,0){4}}
\put(93,8){\vector(0,1){4}}
\put(93,8){\circle{2}}
\put(93,8){\circle*{0.5}}
\put (96.5,9.) {\scriptsize $y$}
\put (93.,12.5) {\scriptsize $z$}
\put (91,9) {\scriptsize $x$}
\end{overpic}\vspace{2mm}\\
\begin{overpic}[width=\columnwidth]{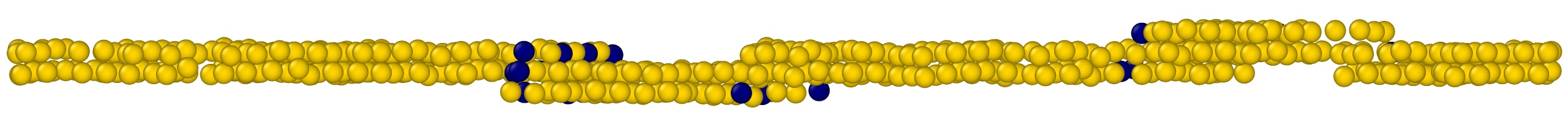}\put (1,7) {\small (b)}
\end{overpic}\vspace{1mm}\\
\begin{overpic}[width=\columnwidth]{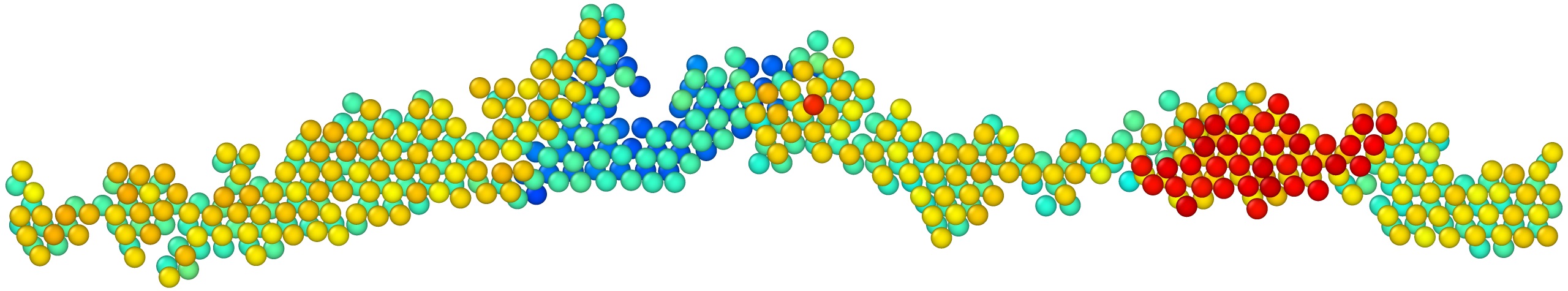}\put (1,11) {\small (c)}
\put(93,13){\vector(1,0){4}}
\put(93,13){\vector(0,1){4}}
\put(93,13){\circle{2}}
\put(93,13){\circle*{0.5}}
\put (96.5,14.) {\scriptsize $y$}
\put (93.,17.5) {\scriptsize $x$}
\put (91,14) {\scriptsize $z$}
\end{overpic}\vspace{0.2mm}\\
\begin{overpic}[width=\columnwidth]{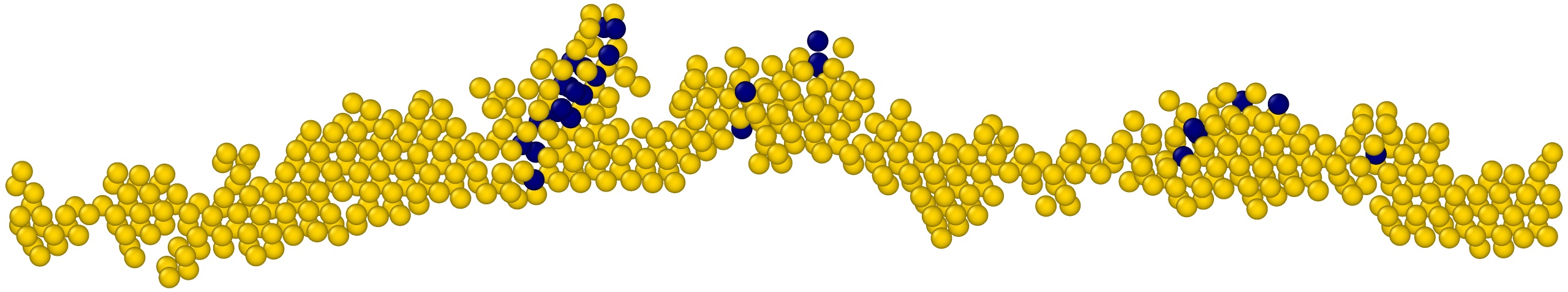}\put (1,11) {\small (d)}\end{overpic}
\caption{Dislocation 1 at $t = 8.50$ ps shown from different views, colored according to (a,c) the $z$ coordinate and (b,d) Voronoi topology.\label{dislocations}}
\end{center}
\end{figure}

To illustrate the manner in which Voronoi topology can also aid in identifying kinks in the dislocation, Figure \ref{dislocations} provides closer views of Dislocation 1 at time $t = 8.50$ ps.  Atoms in subfigures (b) and (d) are colored according to Voronoi topology.  Atoms with FCC local structure are not shown; yellow atoms are those with HCP local structure.  
Dark blue atoms are those whose Voronoi topologies belong neither to the FCC nor HCP families; these atoms belong to dislocation kinks.  The splitting of each dislocation into two partials, separated by a thin ribbon-like stacking fault, is also readily apparent.  In particular, the location and width of each stacking fault can be easily determined.  The clear atomic-scale picture provided by Voronoi topology in this high-temperature system enables the automated analysis of dislocation behavior in a manner not possible using conventional methods.  Figures \ref{screws} and \ref{dislocations}, and the online video, were created using the additional analyses described in Sections \ref{indeterminate} and \ref{clusteranalysis}.

\section{{\it VoroTop} software}
\label{software}

{\it VoroTop} was designed to automate the analysis described in Sections \ref{basics} and \ref{extensions}, and make it accessible to a diverse population of users.  
 
\subsection{Language, license, and availability}
{\it VoroTop} is written in {\tt C++11} to make it compatible with all major operating systems, and is released under an OpenSource BSD 3-Clause license. This license permits redistribution and use of source and binaries, with or without modification, to both academic and for-profit groups.  Source code for the current version (0.4.0 as of publication) is available online in a git repository available at https://gitlab.com/mLazar/VoroTop/.

To make Voronoi topology analysis more broadly accessible, basic features of {\it VoroTop} have also been integrated into the OVITO software package.  OVITO is a scientific visualization and analysis software package for atomistic simulation data developed by Alexander Stukowski \cite{stukowski2009visualization}.  The program has been released as Open Source under the GNU General Public License, and is available for all major platforms.

\subsection{Filters}
\label{filters}

A {\it filter} is an enumeration of one or more families of topologies.  Filters enable {\it VoroTop} to identify crystalline and defect structure.  For example, a filter enumerating all topologies belonging to both the FCC  and HCP families can be used to analyze polycrystalline aluminum.  Such systems consist primarily of crystals with FCC local structure, but also contains defects such as stacking faults, whose atoms have HCP local structure.  

{\bf File format.}  A filter file is divided into three parts.  Lines in the first part begin with a `{\tt \#}' and are used for comments.  Lines in the second part begin with a `{\tt *}' and specify structure types.  Each such line, after the `{\tt *}', includes an index and a plain-text name for the associated structure type.  An indeterminate structure type can be indicated by including two additional indices after the name; these indices indicate the structure types to which this indeterminate type can resolve under perturbations.  Indices of structure types are listed in increasing order and begin with 1.  Remaining lines record Voronoi topologies and their associated structure types, one topology at a time.  Each line begins with a structure type index, and is followed by an integer sequence called a {\it Weinberg vector} that completely describes a Voronoi cell topology; the format of the Weinberg vectors is that found in the work of Weinberg \cite{1966weinberg1}.  

Filter files for several common structure types, included those considered in this paper, are available for download at www.vorotop.org.

\subsection{Performance, optimization, and runtime}

After computing the Voronoi cells using {\tt Voro++}, the calculations of Weinberg vectors scales linearly with the number of atoms in the system.  Running on a single 2.53GHz processor, {\it VoroTop} can currently compute about 30,000 Weinberg vectors per second, or roughly two million atoms per minute.  Parallel processing is not currently implemented in {\it VoroTop}.  The analysis described in Section \ref{indeterminate} to resolve indeterminate topologies also scales linearly in the number of atoms, and in the number of random perturbations considered.

\subsection{Command-line options}
\label{options}

Features of the {\it VoroTop} program are controlled through command-line options.  {\it VoroTop} begins by reading in atomistic data; the LAMMPS dump \cite{plimpton1995fast} and AtomEye extended cfg \cite{li2003atomeye} formats are currently supported.  If specified, a filter file is also read.  Next, the {\tt Voro++} library \cite{2009rycroft} is used to compute the Voronoi cell of each atom, and an algorithm of Weinberg \cite{1966weinberg1} is used to compute Voronoi cell topologies.  Finally, user-specified analyses are performed and output is saved to disk; all output is saved in plain-text format.  

\subsubsection*{{\tt -f}\hspace{4mm} load filter file}
Specifies a filter file to use for analysis.  If this option is used, then a new output file will be created that includes the original data plus the structure types as determined by the given filter.

\subsubsection*{{\tt -w}\hspace{4mm} Weinberg vectors} 
Weinberg's graph-tracing algorithm \cite{1966weinberg1} is used to compute the topology of each Voronoi cell and store it as a vector of integers, which is recorded to output.  The following is also recorded for each atom: the number of faces, the face-index (i.e., the number of faces with each number of edges), and the order of the automorphism group \cite{1966weinberg2} of the Voronoi cell.  The chirality of the Voronoi cell ($+1$ or $-1$ for chiral cells, 0 for achiral ones) is also recorded. 

\subsubsection*{{\tt -d}\hspace{4mm} distribution of Weinberg vectors} 
This option calculates the distribution of Voronoi topologies, and records it using Weinberg vectors.  
 
\subsubsection*{{\tt -g}\hspace{4mm} distribution in perturbed system} 
This feature implements the Monte Carlo analysis described in Section \ref{filterdev}.  In particular, it computes a distribution of topologies from perturbed versions of the input system.  This option requires specifying the number of samples and the desired magnitude of the perturbation.  Perturbations are sampled from a normal Gaussian centered at the origin and with a standard deviation equal to the specified magnitude times the cubed root of the Voronoi cell volume; a magnitude of 0.05 is suggested.

\subsubsection*{{\tt -r}\hspace{4mm} resolve indeterminate topologies} 
This feature implements the analysis described in Section \ref{indeterminate}.  In particular, to each indeterminate structure type is associated two resolving structure types, with one preferred.  If more perturbed copies resolve to the non-preferred structure type than to the preferred type, then that is considered the resolved type; otherwise the preferred type is considered the resolved type.  This feature requires a filter that includes information about which structure types are indeterminate and how they should be resolved.  By default, 5 perturbations are considered, though a different number can be specified at the command line.  Output includes both the indeterminate and resolved structure types.  
 
\subsubsection*{{\tt -c}\hspace{4mm} cluster analysis} 
This feature implements the cluster analysis described in Section \ref{clusteranalysis}.  Each defect and crystal cluster is assigned a unique index, ordered by size.  Positive indices indicate crystal clusters; negative indices indicate defect clusters.  Also recorded for each atom is the size of the cluster to which it belongs.  

By default, all atoms with structure types listed in the specified filter are treated as crystalline, and defect clusters are built only from atoms whose structure types are not listed.  If an index is specified, then only atoms with that structure type are considered crystalline for the purpose of cluster analysis, and clusters are built from atoms with other structure types.  If the {\tt -r} option is specified, then cluster analysis uses resolved types for analysis. 

\subsubsection*{{\tt -od, -o}\hspace{4mm} change output file directory, base name}
By default, {\it VoroTop} saves output to the directory in which the input data is located, and using the input filename as a base name.  For example, if the input filename is {\tt \textasciitilde/research/SFT.00100}, then Weinberg vectors will be saved to {\tt \textasciitilde/research/SFT.00100.wvectors} and a distribution of Weinberg vectors will be saved to {\tt \textasciitilde/research/SFT.00100.distribution}.  The {\tt -od} option allows specification of an alternate output directory; the {\tt -o} options allows specification of an alternate output base name.

\subsection{Examples}
\label{examples}

We present two examples that illustrate how {\it VoroTop} can be used. 

\subsubsection*{Stacking-fault tetrahedron}
Figures \ref{indet}(a-c) in Section \ref{indeterminate} were created using a filter that includes three structure types: determinate FCC, determinate HCP, and indeterminate FCC-HCP.  Analysis is performed on an input file named {\tt SFT} and with a filter named {\tt FCC-both-HCP.filter} using the following commands: 
\vspace{2mm}\\
\indent (a) {\tt ./VoroTop SFT -f FCC-both-HCP.filter}\\
\indent (b) {\tt ./VoroTop SFT -f FCC-both-HCP.filter -r 0}\\
\indent (c) {\tt ./VoroTop SFT -f FCC-both-HCP.filter -r 5}
\vspace{2mm}\\
\noindent Command (a) identifies three local structure types: determinate FCC, indeterminate FCC-HCP, and determinate HCP, each with a separate structure type index.  The {\tt -r} option is not used, so only the initial structure type is recorded; atoms with Voronoi topologies not belonging to this filter are identified as defects using index 0.  

Command (b) uses the {\tt -r} option, but does not run the analysis (or runs it zero times).  The condition for resolving to HCP local structure, as specified in this particular filter, is that more perturbations result in HCP than in FCC local structure.  When no perturbations are considered, all indeterminate types resolve to the default FCC type.  

Command (c) uses the {\tt -r} option with 5 perturbations.  Atoms whose Voronoi topologies resolve more often to HCP than to FCC are considered as having local HCP structure; atoms whose Voronoi topologies resolve as often or more often to FCC than to HCP are considered as having local FCC structure.  

Since the input file {\tt SFT} is formatted as a LAMMPS dump file, each of these commands results in a new file called {\tt SFT.lammps}.  The output file contains the original data, as well as both the initial and resolved structure types.

\subsubsection*{Grain boundary characterization}

In addition to identifying defects, {\it VoroTop} can also be used to characterize them.  We consider one relevant example.  At 0 K, a $\Sigma 5$ [001] (310) symmetric tilt boundary in BCC tungsten has three metastable phases \cite{lazar2015topological}.  The ``structure'' of these phases can be made precise by considering them as repeating patterns of Voronoi topologies.  To determine these patterns, we first use {\it VoroTop} to compute the Weinberg vectors for each atom in the three phases. 
\vspace{2mm}\\
\indent {\tt ./VoroTop phase1 -w}\\
\indent {\tt ./VoroTop phase2 -w}\\
\indent {\tt ./VoroTop phase3 -w}
\vspace{2mm}\\
\noindent Since {\it VoroTop} cannot currently automate the creation of filters, a plain-text editor is used to identify Weinberg vectors of atoms located near the three grain boundaries.  Apart from the single Voronoi topology belonging to the BCC family, atoms near a Phase I grain boundary have three other Voronoi topologies, atoms near a Phase II grain boundary have two other Voronoi topologies, and atoms near a Phase III grain boundary have six other Voronoi topologies; there is no overlap among these sets.  These Weinberg vectors are then used to create a filter which we call {\tt B123.filter}, and which we use to identify grain boundary structures in our samples.  To make the illustrations more clear, we assigned different structure types to each Voronoi topology, so that we can highlight distinct topologies and repeating patterns of them in the three samples.  

\begin{figure}
\begin{center}
\begin{tabular}[c]{ccc}
\fbox{\includegraphics[width=0.25\columnwidth,trim={0.1cm 8.5cm 0.1cm 8.5cm},clip]{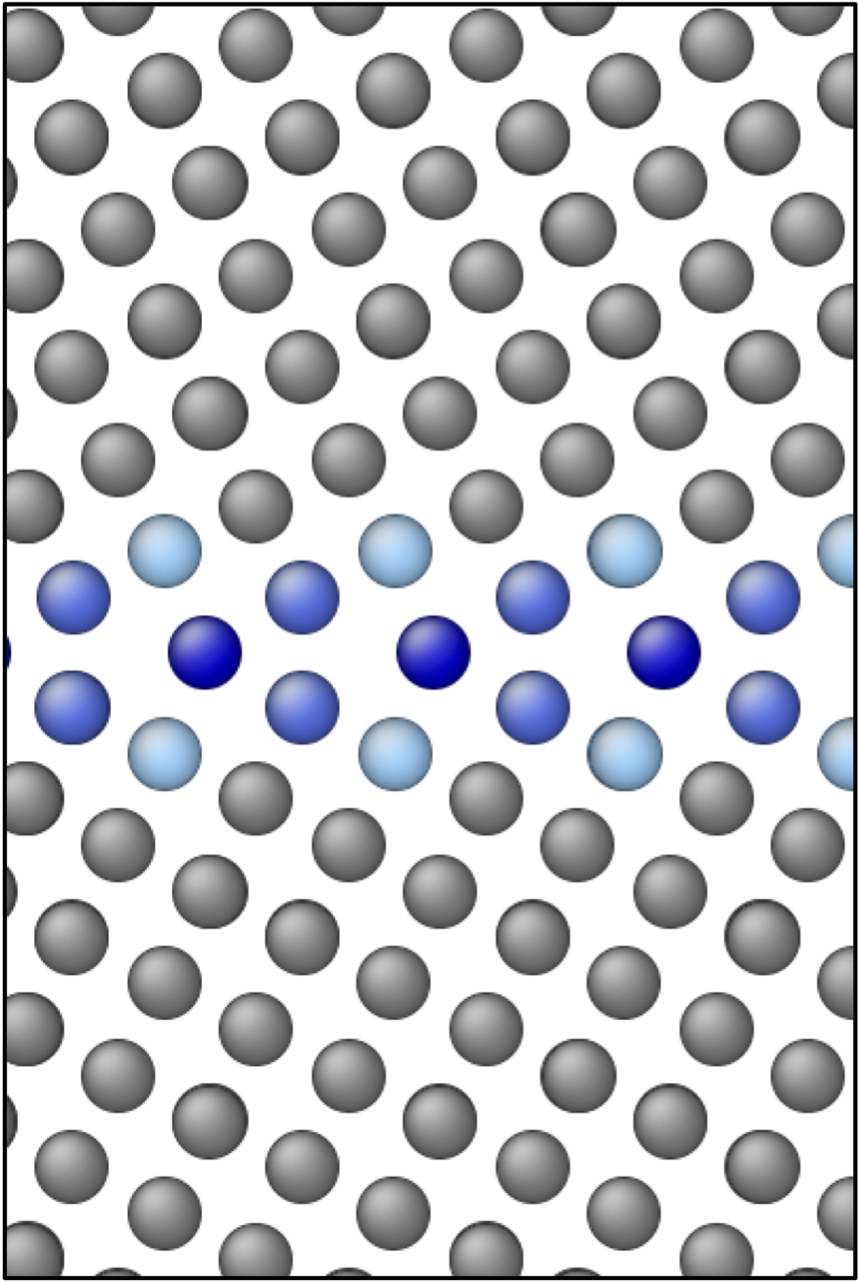}}  & 
\fbox{\includegraphics[width=0.25\columnwidth,trim={0.1cm 8.5cm 0.1cm 8.5cm},clip]{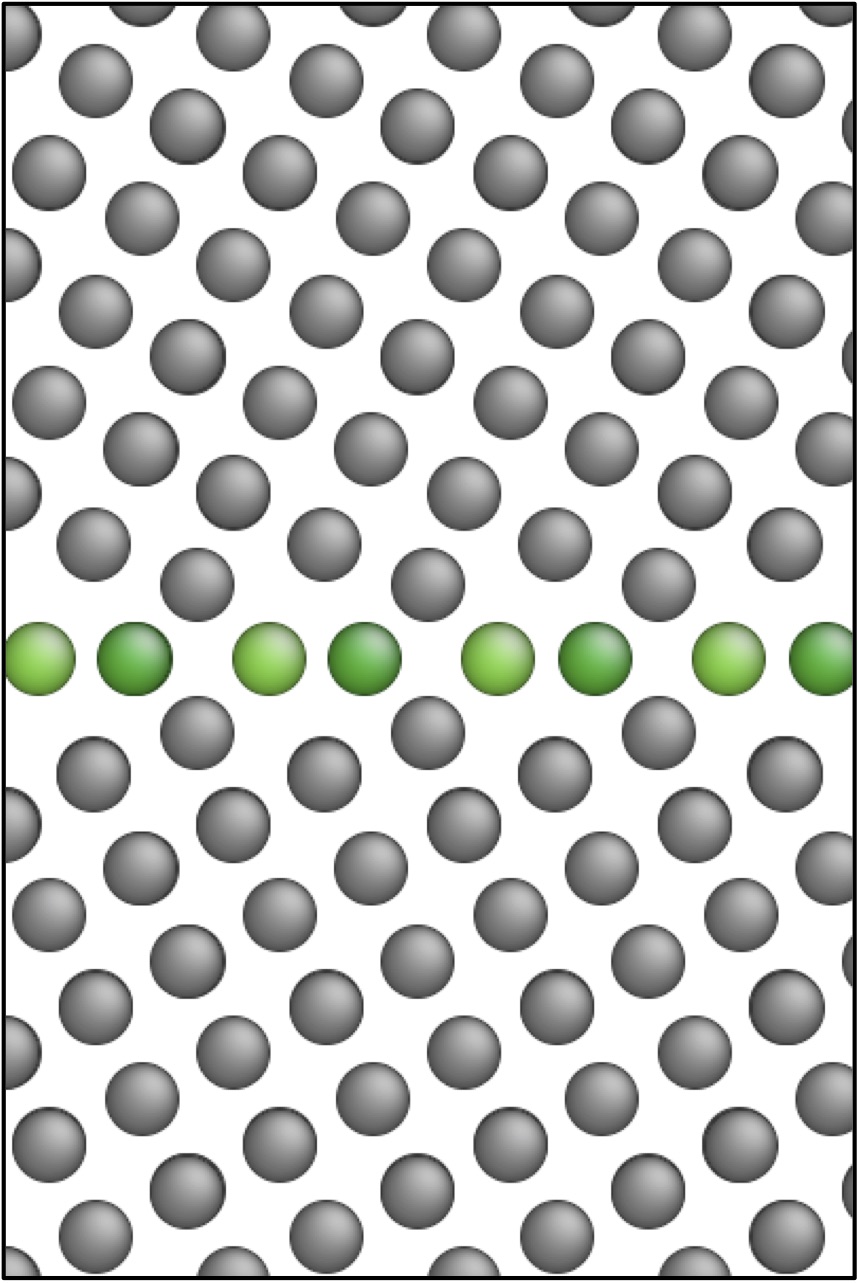}}  & 
\fbox{\includegraphics[width=0.25\columnwidth,trim={0.1cm 8.5cm 0.1cm 8.5cm},clip]{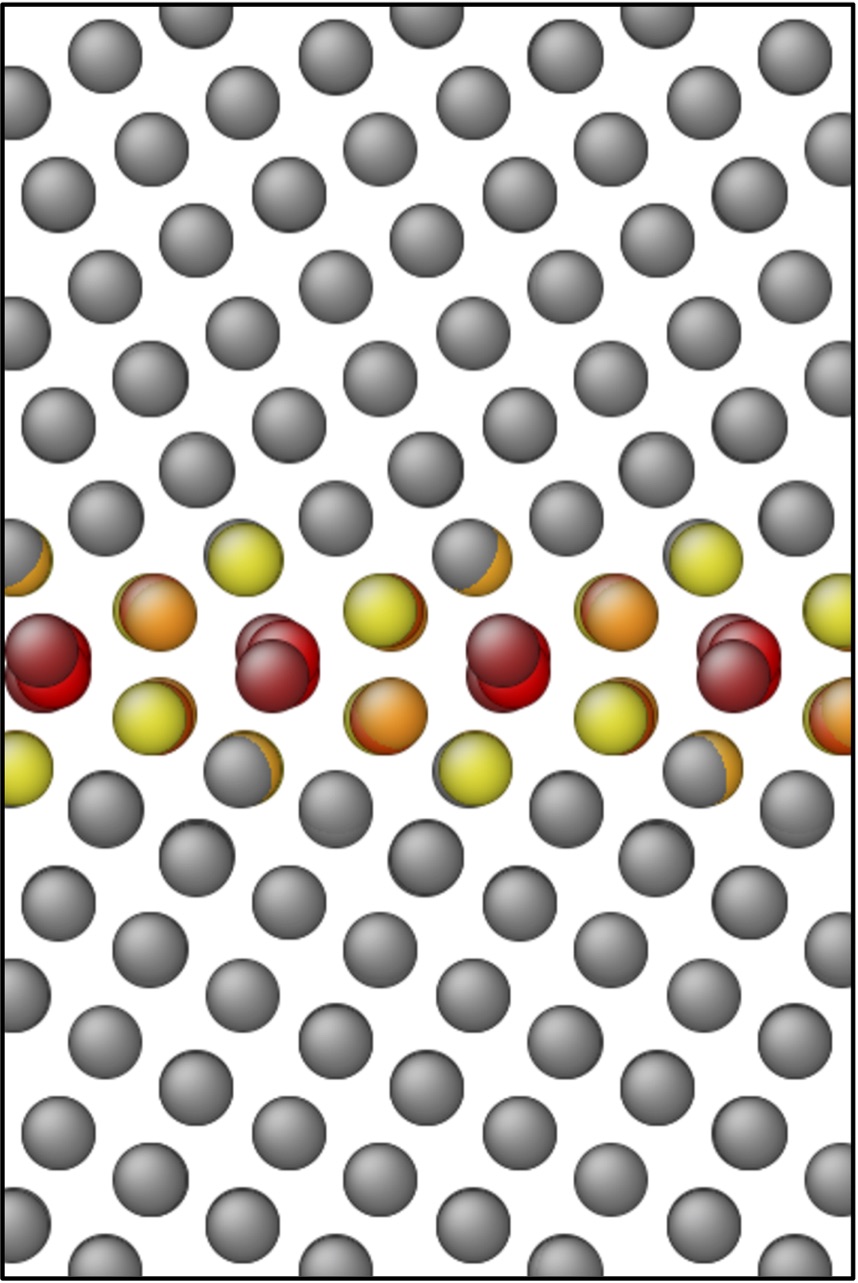}} \\  
\fbox{\includegraphics[width=0.25\columnwidth,trim={0 4cm 0 4cm},clip]{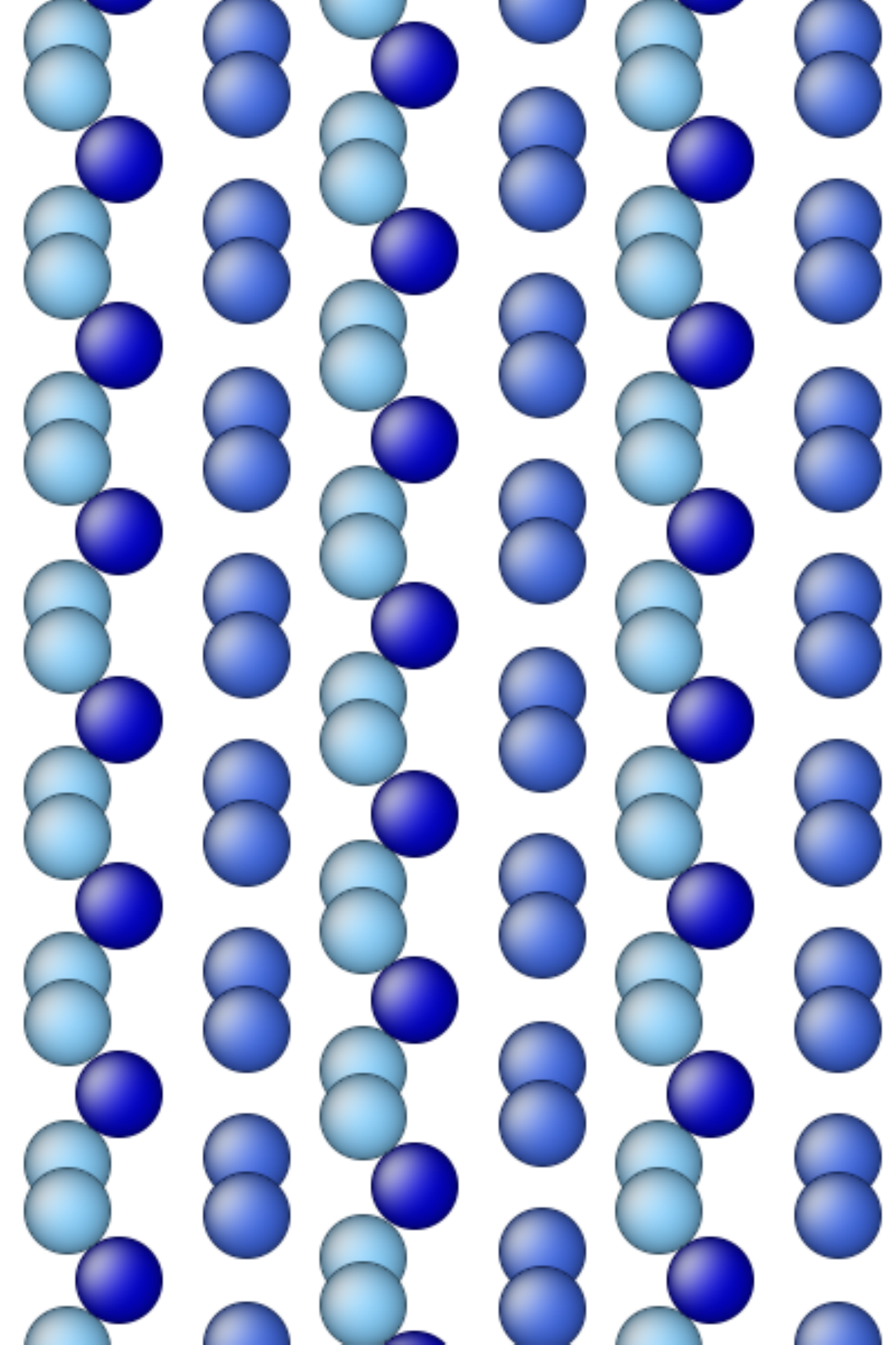}}  & 
\fbox{\includegraphics[width=0.25\columnwidth,trim={0 4cm 0 4cm},clip]{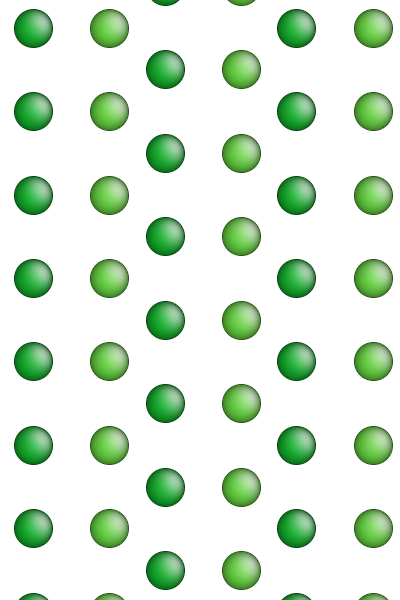}}  &
\fbox{\includegraphics[width=0.25\columnwidth,trim={0 4cm 0 4cm},clip]{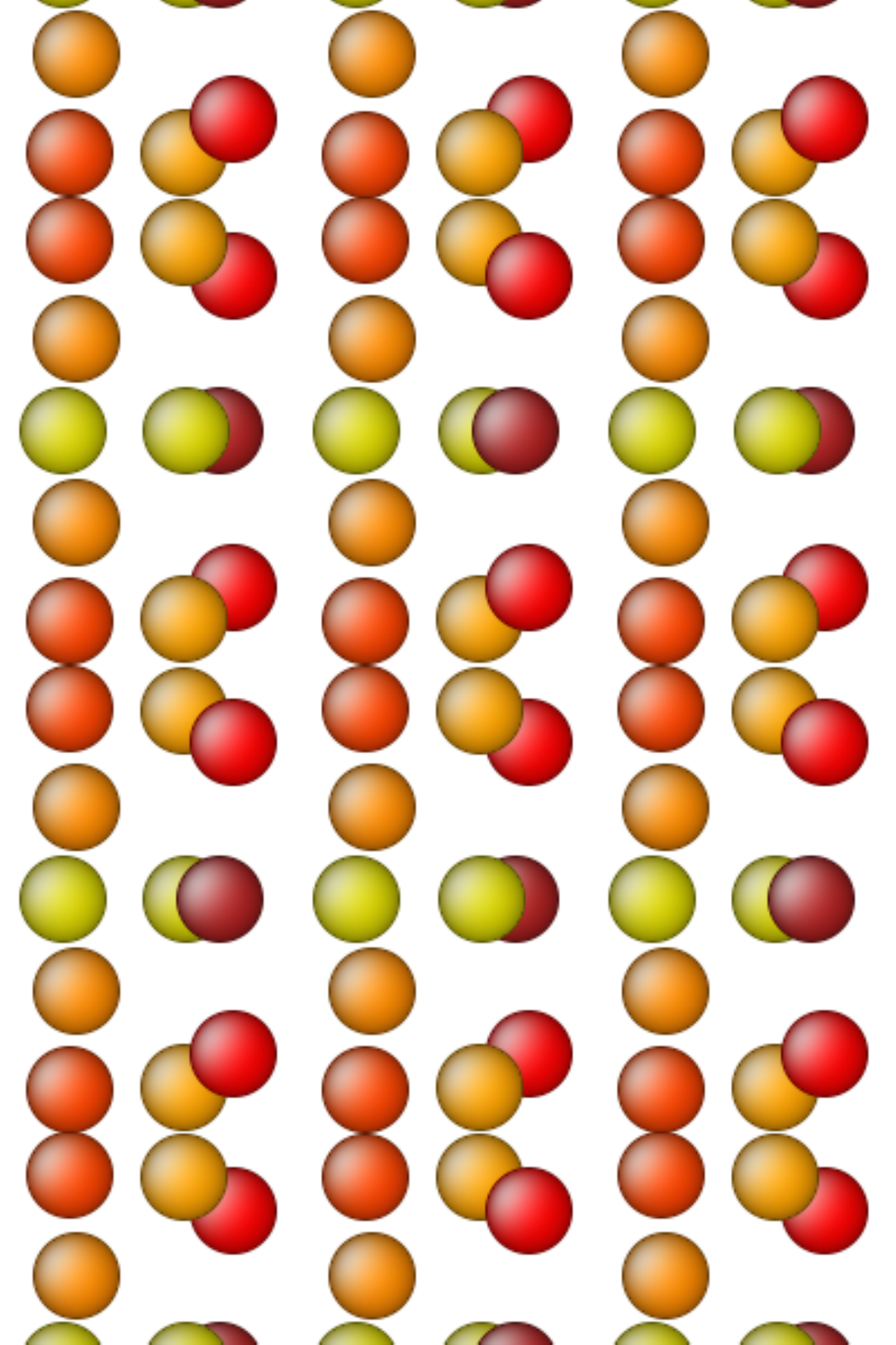}} \\  
(a) Phase I & (b) Phase II & (c) Phase III \\
\end{tabular}
\end{center}
\caption{Profile and planar views of three meta-stable phases of a $\Sigma 5$ [001] (310) symmetric tilt boundary in BCC tungsten.  Each Voronoi topology is assigned a distinct color.  In the profile view, atoms with BCC topologies are shown in grey; in the planar view, these atoms are not shown. \label{sigma5}}
\end{figure}

The following commands are then used to re-analyze the three samples:
\vspace{2mm}\\
\indent {\tt ./VoroTop phase1 -f B123.filter}\\
\indent {\tt ./VoroTop phase2 -f B123.filter}\\
\indent {\tt ./VoroTop phase3 -f B123.filter}
\vspace{2mm}\\
\noindent 
Figure \ref{sigma5} shows profile and planar views of the results of this analysis; each color indicates a distinct Voronoi topology.   
\setlength{\fboxsep}{0pt}

\begin{figure}
\begin{center}
\fbox{\includegraphics[width=0.8\columnwidth,trim={9.cm 0 4.5cm 5cm},clip]{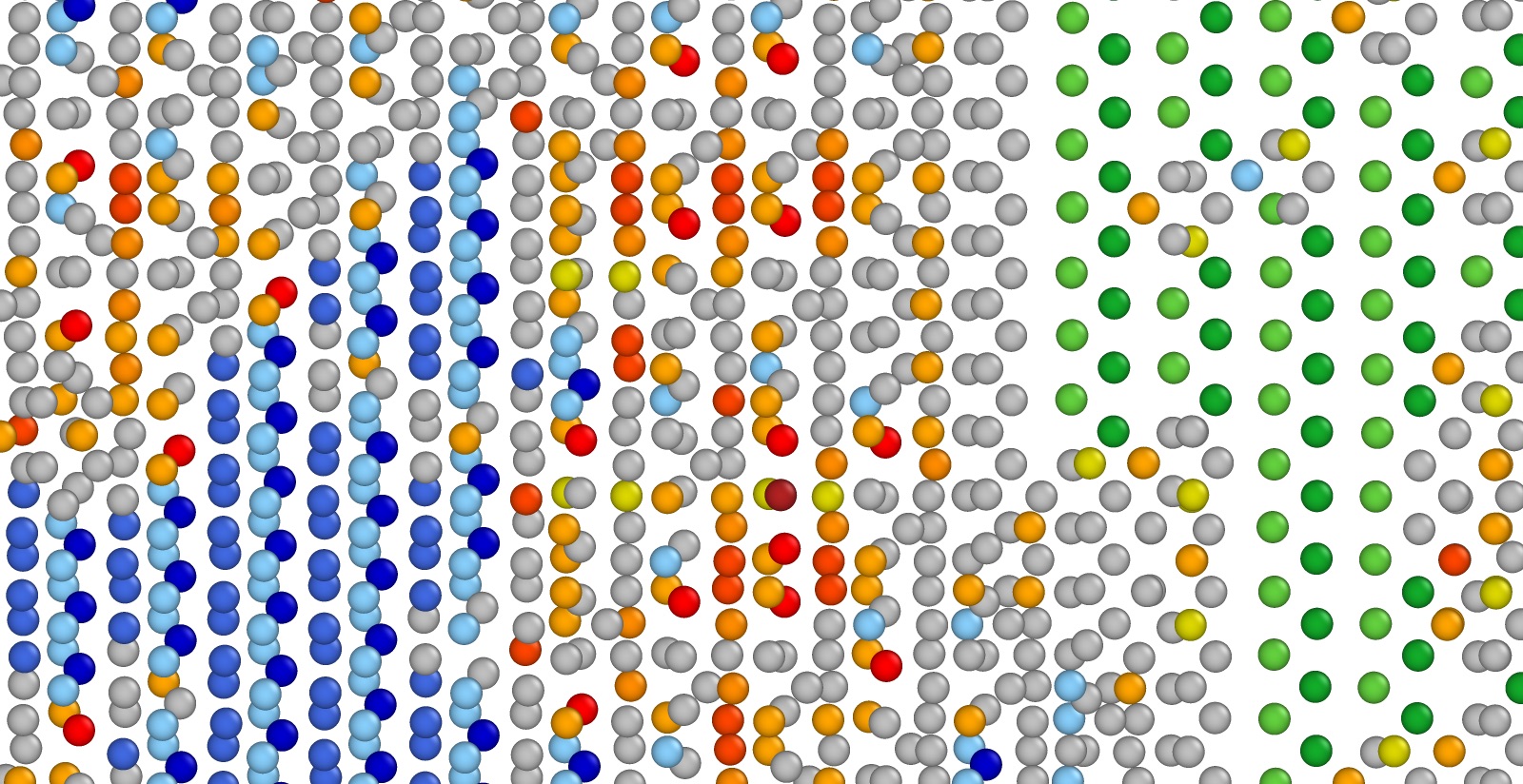}}  
\caption{Grain boundary after irradiation; atoms colored by Voronoi topology.\label{sigma5b}}
\end{center}
\end{figure}

The following command is then used to analyze a more realistic system:  
\vspace{2mm}\\
\indent {\tt ./VoroTop time.0050000 -f B123.filter}
\vspace{2mm}\\
\noindent This grain boundary was initially constructed in Phase I of BCC tungsten and equilibrated at 1500 K, or roughly 40\% of its bulk melting temperature.  Self-interstitial atoms were randomly inserted at the grain boundary to mimic the effects of radiation damage.  The insertion of these atoms transformed the grain boundary from Phase I to a mixture of the three phases.  The results of this analysis can be observed in Figure \ref{sigma5b}.  This characterization of grain boundary structure enables further automated analysis \cite{lazar2015topological}.  

{\bf Acknowledgments.} Development of {\it VoroTop} has been made possible through the generous support of the NSF Division of Materials Research through Award 1507013.  Thanks to Jian Han, Thomas Spencer, and David Srolovitz for many interesting discussions.

\bibliographystyle{ieeetr} 

\end{document}